\documentclass[a4paper,11pt]{article}
\usepackage{jheppub}

\usepackage{graphicx}
\usepackage{subfigure}
\usepackage{pstricks}

\usepackage{amsmath}
\usepackage{array}
\usepackage{MnSymbol}
\usepackage{dsfont}
\usepackage{bbold}
\usepackage{slashed}
\usepackage{mathtools}
\usepackage{mathrsfs}

\usepackage[squaren,Gray]{SIunits}

\usepackage{booktabs}

\newcommand{\cheven}{\text{even}}
\newcommand{\chodd}{\text{odd}}
\newcommand{\tr}[1]{\operatorname{tr} \left\{ #1 \right\}}

\newcommand{\DeltaHone}{\operatorname{\Delta H_1}}
\newcommand{\DeltaHtwo}{\operatorname{\Delta H_2}}
\newcommand{\DeltaHthree}{\operatorname{\Delta H_3}}

\newcommand{\overbar}[1]{\mkern 5mu\overline{\mkern-5mu#1\mkern-2mu}\mkern 2mu}

\newcolumntype{K}{>{\centering\arraybackslash$}p{0.6cm}<{$}}
\renewcommand{\tabcolsep}{0.6cm}

\allowdisplaybreaks

\title{Model-independent calculation of \boldmath $\operatorname{SU}(3)_f$ violation in baryon octet light-cone distribution amplitudes}
\author[]{Philipp Wein,}
\author[]{Andreas Sch\"afer}
\emailAdd{philipp.wein@physik.uni-regensburg.de}
\emailAdd{andreas.schaefer@physik.uni-regensburg.de}
\affiliation[]{Institut f{\"u}r Theoretische Physik, Universit{\"a}t Regensburg,\\
Universit{\"a}tsstra{\ss}e 31, D-93040 Regensburg, Germany}
\date{\today}

\abstract{In this work we present a minimal parametrization of the light-cone distribution amplitudes of the baryon octet including higher twist contributions. Simultaneously we obtain the quark mass dependence of the amplitudes at leading one-loop accuracy by the use of three-flavor baryon chiral perturbation theory (BChPT), which automatically yields model-independent results for the leading $\operatorname{SU}(3)$ flavor breaking effects. For that purpose we have constructed the nonlocal light-cone three-quark operators in terms of baryon octet and meson fields and have carried out a next-to-leading order BChPT calculation. We were able to find a minimal set of distribution amplitudes (DAs) that do not mix under chiral extrapolation towards the physical point and naturally embed the $\Lambda$ baryon. Additionally they are chosen in such a way that all DAs of a certain symmetry class have a similar quark mass dependence (independent of the twist of the corresponding amplitude), which allows for a compact presentation. The results are well-suited for the extrapolation of lattice data and for model building.}


\arxivnumber{1501.07218}

\begin{document}

\maketitle

\flushbottom
\section{Introduction}
Due to the unstable nature of the weakly decaying hyperons there are no scattering experiments with hyperons in the initial state. However, they naturally occur in the final state, for instance in baryon-antibaryon pair production via electron-positron annihilation $e^+e^-\longrightarrow \bar B B$, in deeply virtual exclusive meson electroproduction $\gamma^* p \longrightarrow K^+ \Lambda,\ K^+ \Sigma^0,\ K^0 \Sigma^+$, and in decays of heavy quarkonia to baryon-antibaryon pairs like $J/\Psi,\ \Upsilon \longrightarrow \bar BB$. The standard way to parametrize the nonperturbative information contained in such exclusive processes are (transition) generalized parton distributions or ordinary form factors. At high momentum transfer the contributions from Fock states containing more than the minimal number of partons are power-suppressed and the process can be approximated by a convolution of the involved distribution amplitudes (DAs) with the process-dependent hard scattering kernel. The requirement of large momentum transfer, the instability of the final state hadrons and the fact that distribution amplitudes only occur in convolutions require high luminosity and high granularity detectors to extract information on the hyperon DAs from experiment.\par%
Another type of process where hyperon DAs are involved are the exclusive rare decays of $b$-baryons, like $\Xi_b$, $\Lambda_b$, $\Sigma_b$ and $\Omega_b$, into octet baryons (plus $\gamma$, $l^+ l^-$, \dots). Due to the large mass difference one can hope that higher order Fock states are sufficiently suppressed to allow for a description by three-quark DAs. Since the bottom baryons are produced with increasing rates at LHC and at B-factories worldwide, we have to expect that ever more precise experimental results will be available in future, even for rare decays containing flavor-changing neutral currents, which are sensitive to new physics. Notwithstanding the fact that $b$-baryons are produced at much lower rates than $b$-mesons, they are not less interesting since they allow for an examination of the helicity structure of the $b \longrightarrow s$ transition and thus complement the measurements in the meson sector~\cite{Aaij:2013qta}. As shown in refs.~\cite{Aliev:2002tr,Aliev:2002ww} there are possible scenarios where deviations from the standard model are not seen in the branching ratio of $\Lambda_b\longrightarrow\Lambda l^+l^-$ but only in the $\Lambda$ baryon polarization. It is therefore mandatory to establish a theoretical basis for the description of such decays, and the knowledge of hyperon DAs is one important ingredient. Even the higher twist components can yield relevant contributions~\cite{Wang:2008sm}. Note that constraining the shape of wave functions by calculating the moments of the DAs with lattice QCD plays an even more important role for hyperons than for nucleons, since experimental bounds are less strict than in the nucleon sector.\par%
A first parametrization of the leading twist contributions in hyperon wave functions was already presented in ref.~\cite{Chernyak:1987nu}. A complete parametrization (including all contributions from higher twist) of baryon-to-vacuum matrix elements was first performed for the case of the nucleon in ref.~\cite{Braun:2000kw}, where it turned out that higher twist contributions can yield substantial effects in the baryon sector, since the corresponding normalization constants $\lambda^N_1$ and $\lambda^N_2$ are large compared to the leading twist wave function normalization constant $f^N$. The same procedure has later on been reused in refs.~\cite{Liu:2008yg,Liu:2009uc} to give similar parametrizations for matrix elements of the hyperons in the baryon octet, namely $\Sigma^\pm$, $\Sigma^0$, $\Xi^-$, $\Xi^0$ and $\Lambda$. Our work unifies these different approaches and we find relations between the distribution amplitudes for different baryons even if $\operatorname{SU}(3)_f$ symmetry is broken. The obtained relations are exact including terms up to first order in the quark masses. In this sense we call our results model-independent. However, one should keep in mind that higher order contributions which lie beyond the accuracy of our analysis are model-dependent indeed, since they are affected by the neglection of higher order terms during operator construction and by the choice of the regularization scheme.\par%
As shown in refs.~\cite{Liu:2008yg,Liu:2009uc} one has to introduce six additional DAs if one extends the formalism from the nucleon doublet to the complete baryon octet. Our results show that these additional DAs are determined by the eight independent DAs already known from the nucleon sector. I.e., if one knows the eight standard DAs (and their dependence on the mass splitting between light and strange quarks) for the $\Lambda$ and for at least two types of octet baryons with nonzero isospin, one can predict all the rest. Using the parametrization given in refs.~\cite{Anikin:2013aka,Anikin:2013yoa,Braun:2008ia}, where contributions of Wandzura-Wilczek type~\cite{Wandzura:1977qf} are taken into account, and applying the approximation advocated in ref.~\cite{Anikin:2013aka}, where contributions that can mix with four-particle operators are systematically neglected, we need only $43$ parameters to describe the complete set of baryon octet DAs, including their dependence on the splitting between light and strange quark mass. For details see section~\ref{sect_example_of_application}. This amounts to a significant reduction of parameters compared to an ad~hoc linear extrapolation without the knowledge of $\operatorname{SU}(3)_f$ symmetry breaking, which would require $72$ parameters for the given setup. Therefore our results are useful for the extrapolation of lattice data. In a first step it can be checked whether the nontrivial relations between the different DAs that we have obtained are realised in lattice simulations. If this is the case to a satisfactory degree, one can perform a simultaneous fit to all DAs, which, owing to the significant reduction of parameters mentioned above, has much higher precision. Note that the parameters occurring in the approximation described above are determined by the zeroth, first and second moments of the leading twist DAs and by the zeroth and first moments of twist $4$ DAs, which are, apart from the first moments of the higher twist amplitudes, within reach of state of the art lattice simulations (see ref.~\cite{PhysRevD.89.094511}).\par%
Let us note that $\operatorname{SU}(3)_f$ breaking effects can be of considerable strength: e.g.\ sum rule estimations of the symmetry breaking for leading twist wave function normalization constants range from $\sim 10 \%$, see ref.~\cite{Chernyak:1987nu}, to $\sim 50 \%$. The latter is obtained if one takes the values for $f^\Sigma$ and $f^\Xi$ from refs.~\cite{Liu:2008yg,Liu:2009uc}. In ref.~\cite{Chernyak:1987nu} it is stated that the impact on the shape of the wave function is even larger (at a scale of $\unit{1}{\giga\electronvolt}$). One therefore has to expect substantial effects also at intermediate scales which are relevant for phenomenological computations. We expect lattice QCD calculations to provide quantitative results for $\operatorname{SU}(3)_f$ breaking effects at physical mean quark mass in the near future. These will provide the means to determine the strength of the symmetry breaking terms in our formalism, allowing us to draw more definite conclusions.\par%
At this point we want to highlight a conclusion that can be drawn from our results, which is of conceptual importance and also affects the nucleon sector: we find that the nonanalytic chiral behaviour of moments of DAs does not depend on the twist of the amplitude. Instead, the leading chiral logarithms in the chiral-odd sector are determined by the type of amplitude to which the corresponding moment contributes. The ones occurring in $\Phi^B_{+,i}$ ($\Phi^B_{-,i}$) amplitudes, which will be defined in eq.~\eqref{definition_superior_DAs}, have the same chiral logarithms as $f^B$ ($\lambda_1^B$). The odd moments of the leading twist DA therefore behave like $\lambda_1^B$ instead of $f^B$, which is quite contrary to the intuitional expectation. The shape parameters occurring in ref.~\cite{Anikin:2013aka} can all be assigned uniquely to one of the two classes, which means that the destinction between moments described above is to some extent already present in currently used parametrizations.\par%
This work is organized as follows: In section~\ref{sect_fundamental_definitions} we present some fundamental definitions to lay the base for the parametrization of the nonlocal three-quark operators in terms of baryon and meson fields, which is performed in section~\ref{sect_operator_construction}. A sketch of the leading one-loop baryon chiral perturbation theory (BChPT) calculation is given in section~\ref{sect_calculation}, where we also explain how we have matched our results to the standard DAs given in ref.~\cite{Braun:2000kw}. In section~\ref{sect_result} we present our main results. We provide a definition for DAs that do not mix under chiral extrapolation and naturally embed the $\Lambda$ baryon. The result section is to the most part self-contained such that the reader can skip the details of the derivation at will. We summarize in section~\ref{sect_summary}.
\section{Fundamental definitions}\label{sect_fundamental_definitions}
There exist various possible realizations of chiral symmetry, which all lead to equal results. In the following we only present the definitions we use in this work. For a detailed treatment of the effective field theory framework we refer to~\cite{Weinberg:1978kz,Gasser:1983yg,Gasser:1984gg,Gasser:1987rb,Krause:1990xc,Bernard:2007zu}. The pseudoscalar fields are contained in%
\begin{align}
 u&=\sqrt{U}=\exp \biggl( \frac{i}{2 F_0} \lambda^a \phi^a \biggr) = \exp \biggl( \frac{i}{2 F_0} \phi \biggr) \ ,
\end{align}
where $\lambda^1$, \dots, $\lambda^8$ are Gell-Mann matrices and $F_0$ is the pion decay constant in the three-flavor chiral limit, which corresponds to the convention where $F_\pi = F_0 + \mathcal{O}(m_\pi^2,m_K^2,m_\eta^2) \approx \unit{92}{\mega\electronvolt}$. The matrix $\phi$ can be written in terms of meson fields%
\begin{align}
\phi &= \sqrt{2} 
\begin{pmatrix}
\frac{1}{\sqrt{2}} \pi^0 + \frac{1}{\sqrt{6}} \eta & \pi^+ & K^+ \\
\pi^- & -\frac{1}{\sqrt{2}} \pi^0 + \frac{1}{\sqrt{6}} \eta & K^0 \\
K^- & \overbar{K}^0 & -\frac{2}{\sqrt{6}} \eta 
\end{pmatrix} \ .
\intertext{The $3 \times 3$ matrix $B$ contains the baryon octet:}
\begin{split}
 B &= 
\begin{pmatrix}
\frac{1}{\sqrt{2}} \Sigma^0 + \frac{1}{\sqrt{6}} \Lambda & \Sigma^+ & p \\
\Sigma^- & -\frac{1}{\sqrt{2}} \Sigma^0 + \frac{1}{\sqrt{6}} \Lambda & n \\
\Xi^- & \Xi^0 & -\frac{2}{\sqrt{6}} \Lambda
\end{pmatrix} \\
&\equiv \kappa_p p + \kappa_n n  + \kappa_{\Sigma^-} \Sigma^-  + \kappa_{\Sigma^0} \Sigma^0 + \kappa_{\Sigma^+} \Sigma^+ + \kappa_{\Xi^-} \Xi^-  + \kappa_{\Xi^0} \Xi^0 + \kappa_\Lambda \Lambda \ ,
\end{split}
\end{align}
where the second line defines the matrices $\kappa_B$. Let us from now on use $X\in\{L,R\}$ and, as a convenient notation, $\bar{L}=R$ and $\bar{R}=L$. Where they are not used as an index, $L$ and $R$ are meant to be elements of $\operatorname{SU}(3)_{L/R}$. Defining $u_R = u$ and $u_L = u^\dagger$ the transformation properties of meson and baryon fields under chiral rotations read%
\begin{subequations}
\begin{align}
 u_{X} &\overset{\hat{\chi}}{\longrightarrow}  X u_{X} K^\dagger  = K u_{X} \bar{X}^\dagger  \ , \\
 B &\overset{\hat{\chi}}{\longrightarrow} K B K^\dagger \ , 
\end{align}
\end{subequations}
with the so-called compensator field $K$, which is a common, nonlinear realization of chiral symmetry~\cite{Coleman:1969sm,Gasser:1987rb}. The covariant derivative acting on a baryon field is defined as%
\begin{align}
 D_\mu B  &= \partial_\mu B + [\Gamma_\mu,B] \ ,
\end{align}
where $\Gamma_\mu$ is called the chiral connection and is given by%
\begin{align}
 \Gamma_\mu &= \frac{1}{2} \bigl( u^\dagger \partial_\mu u + u \partial_\mu u^\dagger \bigr) \ .
\end{align}
The chiral vielbein $u_\mu$ and the quark mass insertions $\chi_\pm$ are defined as%
\begin{subequations}
\begin{align}
  u_\mu &= i \bigl( u^\dagger \partial_\mu u - u \partial_\mu u^\dagger \bigr) \ , \\
  \chi_\pm &=  u^\dagger \chi u^\dagger \pm u \chi^\dagger u  \ ,
\end{align}
\end{subequations}
where $\chi=2 B_0 \mathcal{M}$ includes the quark mass matrix, and transform under chiral rotations as follows:%
\begin{subequations}
\begin{align}
u_\mu &\overset{\hat{\chi}}{\longrightarrow}  K u_\mu K^\dagger \ , \\
\chi_\pm &\overset{\hat{\chi}}{\longrightarrow} K \chi_\pm K^\dagger \ .
\end{align}
\end{subequations}%
\begin{table}[t]
\centering%
\begin{tabular}{l r r r r}
\toprule
$\Gamma$ & $\eta^P_\Gamma$ & $\eta^C_\Gamma$ & $\eta^h_\Gamma$ & $\eta^5_\Gamma$ \\ \midrule
$\mathds{1}$ & $1$ & $-1$ & $1$ & $1$\\
$\gamma_5$ & $-1$ & $-1$ & $-1$ & $1$\\
$\gamma_\mu$ & $(-1)_\mu$ & $1$ & $1$ & $-1$\\
$\gamma_\mu \gamma_5$ & $-(-1)_\mu$ & $-1$ & $1$& $-1$ \\
$\sigma_{\mu\nu}$ & $(-1)_\mu (-1)_\nu$ & $1$ & $1$& $1$ \\ \bottomrule
\end{tabular}%
\caption{\label{symmetry_properties_clifford_algebra} The constants $\eta^P_\Gamma$, $\eta^C_\Gamma$, $\eta^h_\Gamma$ and $\eta^5_\Gamma$ characterizing the symmetry properties of the elements of the Clifford algebra, where $(-1)_\mu$ is $1$ for $\mu=0$ and $-1$ for $\mu=1,2,3$.}
\end{table}%
Finally we define for the elements of the Clifford algebra in a unitary representation%
\begin{subequations}
\begin{align}
\Gamma &= \eta^P_\Gamma \gamma_0 \Gamma \gamma_0 \ , \\
\Gamma^T &= \eta^C_\Gamma C \Gamma C \ , \\
\Gamma^\dagger &= \eta^h_\Gamma \gamma_0 \Gamma \gamma_0 \ ,  \\
\Gamma &= \eta^5_\Gamma \gamma_5 \Gamma \gamma_5 \ ,
\end{align}
\end{subequations}
where $C=i\gamma^2 \gamma^0$ is the charge conjugation matrix. The different $\eta$'s are collected in table~\ref{symmetry_properties_clifford_algebra}.%
\section{Operator construction} \label{sect_operator_construction}
In this section we will construct the light-cone ($n$ is a lightlike four-vector) three-quark operator%
\begin{align}
 q^a_\alpha (a_1 n) q^b_\beta (a_2 n) q^c_\gamma (a_3 n) \ ,
\end{align}
in terms of baryon octet and meson octet fields. The antisymmetrization in color indices (which makes the operator a color singlet) and the Wilson lines connecting the quark fields (providing gauge invariance) are not written out explicitly. $a$, $b$, $c$ are flavor and $\alpha$, $\beta$, $\gamma$ Dirac indices. Note that there are many possible parametrizations owing to the freedom of choice one has by neglecting higher order effects. The task is therefore not only to find a parametrization, but to find one that is most convenient for the loop calculation to be performed and can be easily matched to the standard decomposition given in ref.~\cite{Braun:2000kw}. For the parametrization of the nonlocal operator one needs functions, where the moments of the functions play the role of low energy constants (LECs). For the parametrization presented below these functions can be easily matched to standard distribution amplitudes.%
\subsection{Symmetry properties}\label{sect_sym_3qopp}
To perform the construction of an operator within the effective theory we have to know its symmetry properties. To make use of chiral symmetry it is convenient to split the quark fields in left- and right-handed parts%
\begin{align} \label{3q_op_decompRL}
\begin{split}
  q^a_\alpha (a_1 n) q^b_\beta (a_2 n) q^c_\gamma (a_3 n) &=\mathcal O^{abc}_{RR,\alpha\beta\gamma}(a_1,a_2,a_3) + \mathcal O^{abc}_{LL,\alpha\beta\gamma}(a_1,a_2,a_3) \\
  &\quad+ \mathcal O^{abc}_{RL,\alpha\beta\gamma}(a_1,a_2,a_3) + \mathcal O^{abc}_{LR,\alpha\beta\gamma}(a_1,a_2,a_3) \\
  &\quad+ \mathcal O^{cab}_{RL,\gamma\alpha\beta}(a_3,a_1,a_2) + \mathcal O^{cab}_{LR,\gamma\alpha\beta}(a_3,a_1,a_2) \\
  &\quad+ \mathcal O^{bca}_{RL,\beta\gamma\alpha}(a_2,a_3,a_1) + \mathcal O^{bca}_{LR,\beta\gamma\alpha}(a_2,a_3,a_1) \ ,
\end{split}
\end{align}
where the operators $\mathcal O_{XY}$ for $X$, $Y$ $\in \{L,R\}$ are given by%
\begin{align} \label{3qOp_definition_chiral}
 \mathcal O^{abc}_{XY,\alpha\beta\gamma}(a_1,a_2,a_3) = q^a_{X,\alpha}(a_1 n) q^b_{X,\beta}(a_2 n) q^c_{Y,\gamma}(a_3 n) \ ,
\end{align}
where the left-/right-handed quark fields are defined as $q_{L/R}=\gamma_{L/R} \, q$ with the projection matrices $\gamma_{L/R}=(1\mp \gamma_5)/2$. These operators can be characterized by their transformation properties under parity transformation ($\hat p$), charge ($\hat c$) and hermitian ($\dagger$) conjugation and chiral rotations ($\hat \chi$):%
\begin{subequations}
\begin{align}
 \mathcal O^{abc}_{XY,\alpha\beta\gamma}(a_1,a_2,a_3) &\overset{\hat p}{\longrightarrow} (\gamma_0)_{\alpha \alpha^\prime} (\gamma_0)_{\beta \beta^\prime} (\gamma_0)_{\gamma \gamma^\prime} \mathcal O^{abc}_{\bar X \bar Y,\alpha^\prime \beta^\prime \gamma^\prime}(a_1,a_2,a_3) \label{sym_P} \ , \\
 \mathcal O^{abc}_{XY,\alpha\beta\gamma}(a_1,a_2,a_3) &\overset{\mathclap{\hat c \  \dagger \ \hat p}}{\longrightarrow} -C_{\alpha \alpha^\prime} C_{\beta \beta^\prime} C_{\gamma \gamma^\prime} \mathcal O^{abc}_{XY,\alpha^\prime \beta^\prime \gamma^\prime}(a_1,a_2,a_3) \ , \label{sym_hPC} \\
\mathcal O^{abc}_{XY,\alpha\beta\gamma}(a_1,a_2,a_3) &\overset{\hat \chi}{\longrightarrow} X_{a a^\prime} X_{b b^\prime} Y_{c c^\prime} \mathcal O^{a^\prime b^\prime c^\prime}_{XY,\alpha\beta\gamma}(a_1,a_2,a_3) \ ,
\end{align}
\end{subequations}
where in eq.~\eqref{sym_hPC} charge conjugation is performed first. Additionally we know that each operator transforms under a translation in $n$-direction as%
\begin{align}
\begin{split}
\mathcal O^{abc}_{XY,\alpha\beta\gamma}(a_1+\delta a,a_2+\delta a,a_3+\delta a) &= \exp \left\{i \, \delta a \, n \cdot \hat P \right\}\mathcal O^{abc}_{XY,\alpha\beta\gamma}(a_1,a_2,a_3)\exp \left\{-i \, \delta a \, n \cdot \hat P \right\} \ ,
\end{split}
\end{align}
where $\hat P$ is the momentum operator which acts as a generator of translations. Another symmetry of the three-quark operators defined in eq.~\eqref{3qOp_definition_chiral} is the invariance under the exchange of the quark in the first and the second position or even an invariance under exchange of all three quarks in case of the operators containing right-handed or left-handed fields exclusively. On top of this the operator is invariant if one simultaneously rescales $a_i \longrightarrow \lambda a_i$ and $n_\mu \longrightarrow n_\mu/\lambda$, which we will call scaling property.%
\subsection{Low energy operators}
Using the previously defined fields $u_R$ and $u_L$ we can write down the operators, which contribute to baryon-to-vacuum matrix elements of three-quark currents at leading one-loop level and have correct transformation properties under chiral rotations in the following compact form:%
 \begin{align}\label{eff_opp} 
\begin{split}
  \mathcal O^{abc}_{XY,\alpha\beta\gamma}(a_1,a_2,a_3) &= \int [dx] \sum \limits_{i,j} \sum \limits_{k=1}^{k_j} \mathcal{F}^{i,j,k}_{XY}(x_1,x_2,x_3) \, \Gamma^{i,XXY}_{\alpha \beta \gamma \delta} \, B^{j,k,XXY}_{\delta,abc}(z) \ ,
\end{split}
\end{align}
where the correct transformation behaviour under translations in $n$-direction is ensured by $z_\mu= n_\mu \sum x_i a_i$ and the constraint that $x_1+x_2+x_3=1$ in%
\begin{align}
 \int [dx]&= \int dx_1 dx_2 dx_3 \, \delta \bigl(1-\sum x_i\bigr) \ , 
\end{align}
where the integrations run from $0$ to $1$. The $\mathcal{F}$'s are functions of $x_1$, $x_2$, $x_3$ only and $k_j$ is given in table~\ref{symmetry_properties_B}. The $\Gamma$'s are defined as%
\begin{align}
 \Gamma^{i,XYZ}_{\alpha \beta \gamma \delta} &= (\gamma_X \Gamma^i_A \gamma_Y C)_{\alpha \beta} (\gamma_Z \Gamma^i_B (i \slashed{\partial})^{d^m_i})_{\gamma \delta}  (n \cdot \partial)^{d^n_i} \ , \label{definition_Gamma_structure}
\end{align}
\begin{table}[t]
\centering%
\begin{tabular}{c c c c c c c}
\toprule
$i$ &  $\qquad\mathllap{\Gamma^i_A } \otimes \mathrlap{ \Gamma^i_B}\qquad $ & $\eta^h_{\Gamma,i}=\eta^C_{\Gamma,i}$ &   $\eta^{C}_{\Gamma^A,i}$ &   $\eta^{5}_{\Gamma^B,i}$ & $d^m_i$ & $d^n_i$  \\ \midrule
$1$ & $ \mathllap{i  \mathds{1} } \otimes \mathrlap{ \slashed{n}} $    & $-1$  & $-1$ &$-1$ & $2$ & $-1$\\
$2$ & $ \mathllap{\mathds{1} } \otimes \mathrlap{ \mathds{1}} $  & $\phantom{+}1$  & $-1$ &$\phantom{+}1$ & $1$ & $\phantom{-}0$ \\
$3$ & $ \mathllap{\sigma^{\partial n} } \otimes \mathrlap{ \slashed{n}} $   & $\phantom{+}1$  & $\phantom{+}1$ &$-1$& $2$ & $-2$ \\
$4$ & $ \mathllap{\sigma^{\mu n} } \otimes \mathrlap{ \gamma_\mu} $    & $\phantom{+}1$  & $\phantom{+}1$  &$-1$& $2$ & $-1$\\
$5$ & $ \mathllap{\sigma^{\mu \nu} } \otimes \mathrlap{ \sigma_{\mu \nu}} $    & $\phantom{+}1$  & $\phantom{+}1$ &$\phantom{+}1$& $1$& $\phantom{-}0$ \\
$6$ & $ \mathllap{i  \sigma^{\partial n} } \otimes \mathrlap{ \mathds{1}} $  & $-1$  & $\phantom{+}1$ &$\phantom{+}1$& $1$ & $-1$ \\
$7$ & $ \mathllap{\sigma^{\mu \partial} } \otimes \mathrlap{ \sigma_{\mu n}} $   & $\phantom{+}1$ & $\phantom{+}1$ &$\phantom{+}1$& $1$ & $-1$ \\
$8$ & $ \mathllap{\sigma^{\mu \partial} } \otimes \mathrlap{ \gamma_\mu} $   & $\phantom{+}1$  & $\phantom{+}1$ &$-1$& $0$ & $\phantom{-}0$ \\
$9$ & $ \mathllap{\sigma^{\mu n} } \otimes \mathrlap{ \sigma_{\mu n}} $    & $\phantom{+}1$  & $\phantom{+}1$ &$\phantom{+}1$& $3$ & $-2$ \\ \bottomrule
\end{tabular}
\caption{\label{symmetry_properties_Gamma} List of $\Gamma^i_A \otimes \Gamma^i_B $. $\eta^P_{\Gamma,i}=1$ by choice (see comment in the text). We have multiplied structures $1$ and $6$ with a factor of $i$ such that $\eta^h_{\Gamma,i}=\eta^C_{\Gamma,i}$ for all structures and, thus, $\eta^{hPC}_{\Gamma,i}=1$. In cases where four-vectors are used in the place of Lorentz indices the notation means that the corresponding Lorentz index is contracted with the index of the vector; e.g.\ $\sigma^{\partial n}=\sigma^{\mu\nu} \partial_\mu n_\nu$.}
\end{table}%
where $\Gamma^i_A$, $\Gamma^i_B$, $d^m_i$ and $d^n_i$ can be taken from table~\ref{symmetry_properties_Gamma}. The occurring derivatives act on the $B$'s. We have introduced adequate powers of $i \slashed{\partial}$ to have functions $\mathcal{F}$ of mass dimension $2$, which is compatible with the standard mass dimension of distribution amplitudes. Using $i \slashed{\partial}$ (which leads to a factor $m_B$ in the final result) instead of the baryon mass in the chiral limit $m_0$ (which would be the standard choice) has the advantage that it allows for a straightforward matching of our parametrization to the general decomposition given in ref.~\cite{Braun:2000kw} and to refs.~\mbox{\cite{Liu:2008yg,Liu:2009uc}} (see also section~\ref{sect_matching}). The power of $(n \cdot \partial)$ is chosen such that the scaling property is fulfilled. Note that in the chiral-odd sector one can actually write down more structures, which have the form $\Gamma^{i,XYX}_{\beta \gamma \alpha \delta}$ or $\Gamma^{i,YXX}_{\gamma \alpha \beta \delta}$. However, these structures are not independent. They can be rewritten in terms of $\Gamma^{i,XXY}_{ \alpha \beta \gamma \delta}$ using Fierz transformation. In order to reduce the $\Gamma$'s to the minimal set given in table~\ref{symmetry_properties_Gamma} one has to use the identity $\sigma^{\mu\nu} \gamma_5 = \frac{i}{2} \varepsilon^{\mu\nu\rho\sigma}\sigma_{\rho\sigma}$ and the fact that it is sufficient to construct structures of positive parity (see explanation below eq.~\eqref{parity_constraint}). Additionally one has to use that multiplying both structures $\Gamma^i_A$ and $\Gamma^i_B$ with a $\gamma_5$ does not lead to a new, independent structure owing to the projection with $\gamma_{L/R}$ in eq.~\eqref{definition_Gamma_structure}.\par
The $B$'s in eq.~\eqref{eff_opp} are defined as%
\begin{align}
 B^{j,k,XYZ}_{\delta,abc} &= (u_X)_{a a^\prime} (u_Y)_{b b^\prime} (u_Z)_{c c^\prime} B^{j,k}_{\delta,a^\prime b^\prime c^\prime} \ ,
\end{align}%
\begin{table}[t]
\centering%
\begin{tabular}{c c c c c c}
\toprule
$j$ & $B^j_{1,\delta}$ & $B^j_2$ & $B^j_3$ & trace$^j$  & $k_j$  \\ \midrule
$1$ & $B_\delta$ & $\mathds{1}$ & $\mathds{1}$  & $1$ & $3$ \\
$2$ & $B_\delta$ & $\mathds{1}$ & $\mathds{1}$ & $\tr{\chi_+} m_0^{-2} $  & $3$  \\
$3$ & $B_\delta$ & $\tilde\chi_+ m_0^{-2} $ & $\mathds{1}$ & $1$ & $6$ \\ \bottomrule
\end{tabular}
\caption{\label{symmetry_properties_B} In this table we list only terms which contribute to the one-loop calculation of baryon-to-vacuum matrix elements of the operator. $\tilde \chi_+$ is defined as $\chi_+ -\tr{\chi_+}/3$. This is a convenient choice since this combination (in a leading one-loop calculation) vanishes along the symmetric line, where $m_u=m_d=m_s$.}
\end{table}%
where%
\begin{subequations}
\begin{align}
 B^{j,1}_{\delta,a b c} &= (B^j_{1,\delta})_{a a^\prime} (B^j_2)_{b b^\prime} (B^j_3)_{c c^\prime} \varepsilon_{a^\prime b^\prime c^\prime} \times \text{trace$^j$} \ , \\
 B^{j,2}_{\delta,a b c} &= (B^j_3)_{a a^\prime} (B^j_{1,\delta})_{b b^\prime} (B^j_2)_{c c^\prime} \varepsilon_{a^\prime b^\prime c^\prime} \times \text{trace$^j$} \ , \\
 B^{j,3}_{\delta,a b c} &= (B^j_2)_{a a^\prime} (B^j_3)_{b b^\prime} (B^j_{1,\delta})_{c c^\prime} \varepsilon_{a^\prime b^\prime c^\prime} \times \text{trace$^j$} \ , \\
 B^{j,4}_{\delta,a b c} &= (B^j_2)_{a a^\prime} (B^j_{1,\delta})_{b b^\prime} (B^j_3)_{c c^\prime} \varepsilon_{a^\prime b^\prime c^\prime} \times \text{trace$^j$} \ , \\
 B^{j,5}_{\delta,a b c} &= (B^j_{1,\delta})_{a a^\prime} (B^j_3)_{b b^\prime} (B^j_2)_{c c^\prime} \varepsilon_{a^\prime b^\prime c^\prime} \times \text{trace$^j$} \ , \\
 B^{j,6}_{\delta,a b c} &= (B^j_3)_{a a^\prime} (B^j_2)_{b b^\prime} (B^j_{1,\delta})_{c c^\prime} \varepsilon_{a^\prime b^\prime c^\prime} \times \text{trace$^j$} \ .
\end{align}
\end{subequations}
For cases where $B^j_2=B^j_3$ we only use $  B^{j,1}_{\delta,a b c}$, $ B^{j,2}_{\delta,a b c}$ and $ B^{j,3}_{\delta,a b c}$ and thus $k_j=3$. The different possible combinations of $B$'s can be taken from table~\ref{symmetry_properties_B}. All baryon and meson fields which are connected to each other (by a summation over a shared flavor index) have to be at the same spacetime position, owing to the fact that the compensator field $K$ is a local transformation. However, chiral symmetry actually also allows for the possibility that the trace term in $B$ is situated at a different spacetime position as the rest of the operator. We consider this possibility in appendix~\ref{app_no_cov_der} and show that such a parametrization only differs in higher order terms. Note that no structures of the form $[B_\delta,\tilde\chi_+]$, $\{B_\delta,\tilde\chi_+\}$, or $\tr{B_\delta \tilde\chi_+}$ occur in table~\ref{symmetry_properties_B}, since they can be reexpressed in terms of the third structure, which means that we have only one second order structure ($j=3$) that is responsible for $\operatorname{SU}(3)_f$ breaking. Also the operators which describe the behaviour along the $\operatorname{SU}(3)_f$ symmetric line ($j=1,2$) are not linearly independent, but the situation is more complicated in this case: since operators of the same class (i.e.\ same $j$ but different $k$) are related to each other (see eq.~\eqref{lin_dep_j12}) one has to take care that the symmetry properties of the operator under quark exchange are respected. Therefore, we postpone this discussion to section~\ref{sect_elim}. \par
There are no covariant derivatives acting on the baryon field within the $B$'s. In appendix~\ref{app_no_cov_der} we show that they can always be traded for derivatives acting on the whole structure plus higher order contributions, which can be neglected. This fact will turn out to be very convenient for calculating loop contributions, since the derivatives acting on the complete structure do not lead to additional loop momenta in the integrals. \par
The effective operator given in eq.~\eqref{eff_opp} already transforms correctly under chiral rotations and translations along the light-cone vector $n$. It also fulfills the scaling property. The remaining symmetry properties given in section~\ref{sect_sym_3qopp} will now be implemented by constraining the functions $\mathcal{F}$. We consider%
\begin{subequations}
 \begin{align}
 B^{j,k,XYZ}_{\delta,abc} &\overset{\hat p}{\longrightarrow} (\gamma_0)_{\delta \delta^\prime} B^{j,k,\bar{X}\bar{Y}\bar{Z}}_{\delta^\prime,abc}  \ , \label{P_B} \\
 B^{j,k,XYZ}_{\delta,abc}  &\overset{\mathclap{\hat c \  \dagger \ \hat p}}{\longrightarrow} - C_{\delta \delta^\prime}  B^{j,k,XYZ}_{\delta^\prime,abc}  \ , \label{hPC_B}
\end{align}
\end{subequations}
and%
\begin{subequations}
 \begin{align}
 \Gamma^{i,XYZ}_{\alpha \beta \gamma \delta} &\overset{\hat p}{\longrightarrow} -\eta^P_{\Gamma,i} \ (\gamma_0)_{\alpha \alpha^\prime} (\gamma_0)_{\beta \beta^\prime} (\gamma_0)_{\gamma \gamma^\prime} \ \Gamma^{i,\bar{X}\bar{Y}\bar{Z}}_{\alpha^\prime \beta^\prime \gamma^\prime \delta^\prime}\ (\gamma_0)_{\delta^\prime \delta}  \ , \label{P_Gamma} \\
 \begin{split}
\Gamma^{i,XYZ}_{\alpha \beta \gamma \delta} &\overset{\mathclap{\hat c \  \dagger \ \hat p}}{\longrightarrow} - \eta^{hPC}_{\Gamma,i} \  C_{\alpha \alpha^\prime} C_{\beta \beta^\prime} C_{\gamma \gamma^\prime} \ \Gamma^{i,XYZ}_{\alpha^\prime \beta^\prime \gamma^\prime \delta^\prime} \ C_{\delta^\prime \delta} \ . \label{hPC_Gamma}
 \end{split}
\end{align}
\end{subequations}
Eqs.~\eqref{hPC_B} and \eqref{hPC_Gamma} yield (together with eqs.~\eqref{sym_hPC} and \eqref{eff_opp} and since $\eta^{hPC}_{\Gamma}=1$)%
\begin{align} \label{phase_1}
 \bigl(\mathcal{F}^{i,j,k}_{XY}\bigr)^* &= \mathcal{F}^{i,j,k}_{XY} \ , 
\end{align}
which would mean that the $\mathcal{F}$'s are real-valued. However this argument relies on the assumption that one gets no additional phases from charge conjugation of quarks and baryons, which is not necessarily true. If we allow for such additional phases the above equation has to be generalized to%
\begin{align} \label{phase_2}
 \bigl(\mathcal{F}^{i,j,k}_{XY} e^{i \theta}\bigr)^* &= \mathcal{F}^{i,j,k}_{XY} e^{i \theta} \ , 
\end{align}
where we have an additional overall phase which is equal for all distribution amplitudes. However, this additional phase is unphysical and can be dropped. Eqs.~\eqref{P_B} and \eqref{P_Gamma} yield (together with eqs.~\eqref{sym_P} and \eqref{eff_opp} and since $\eta^P_{\Gamma}=1$)%
\begin{align} \label{parity_constraint}
 {\mathcal{F}^{i,j,k}_{\bar X \bar Y}} &= -\mathcal{F}^{i,j,k}_{XY} \ .
\end{align}
Therefore we only have to differentiate between chiral-even $\mathcal{F}^{i,j,k}_{RR}=-\mathcal{F}^{i,j,k}_{LL}\equiv \mathcal{F}^{i,j,k}_{\cheven}$ and chiral-odd $\mathcal{F}^{i,j,k}_{LR}=-\mathcal{F}^{i,j,k}_{RL}\equiv \mathcal{F}^{i,j,k}_{\chodd}$. Notice that we have chosen to only construct structures $\Gamma_A \otimes \Gamma_B$ which have positive parity. The negative parity structures, which one can obtain by multiplying all $\Gamma_B$ with a $\gamma_5$, would lead to the same operators since eq.~\eqref{parity_constraint} then would yield an extra minus sign.%
\subsection{Symmetry under exchange of quark fields}
In this section we use the symmetry of the original three-quark operators under exchange of quark fields with the same handedness to reduce the number of amplitudes. Using the constraint that the operators have to be equal under exchange of the first and the second quark yields%
\begin{subequations} \label{symmetry_constraints_12}
\begin{align}
 &\text{\underline{$j=1,2$:}} & \mathcal{F}^{i,j,1}_{XY}{(x_1,x_2,x_3)} &= -\eta^C_{\Gamma^A,i} \mathcal{F}^{i,j,2}_{XY} {(x_2,x_1,x_3)} \ , \label{symmetry_constraints_12_kj3_1}\\
 & & \mathcal{F}^{i,j,3}_{XY}{(x_1,x_2,x_3)} &= -\eta^C_{\Gamma^A,i} \mathcal{F}^{i,j,3}_{XY} {(x_2,x_1,x_3)} \ , \label{symmetry_constraints_12_kj3_2} \\[0.5cm]
 &\text{\underline{$j=3$:}} & \mathcal{F}^{i,3,1}_{XY}{(x_1,x_2,x_3)} &= -\eta^C_{\Gamma^A,i} \mathcal{F}^{i,3,4}_{XY} {(x_2,x_1,x_3)} \ , \\
 & & \mathcal{F}^{i,3,2}_{XY}{(x_1,x_2,x_3)} &= -\eta^C_{\Gamma^A,i} \mathcal{F}^{i,3,5}_{XY} {(x_2,x_1,x_3)} \ , \\
 & & \mathcal{F}^{i,3,3}_{XY}{(x_1,x_2,x_3)} &= -\eta^C_{\Gamma^A,i} \mathcal{F}^{i,3,6}_{XY} {(x_2,x_1,x_3)} \ .
\end{align}
\end{subequations}
In the chiral-odd sector one now uses these relations to eliminate $\mathcal{F}^{i,j,2}_{XY}$ (if $j=1,2$) and $\mathcal{F}^{i,3,4/5/6}_{XY}$. Additionally we can use that%
\begin{align} \label{chiral_proj_cons_XXY}
  (\gamma_Y \Gamma_A \gamma_X C)_{\gamma \beta} (\gamma_X \Gamma_B)_{\alpha \delta} &= 0 \ , \quad \text{if } X\neq Y \text{ and } \Gamma_A \in \{ \mathds{1}, \gamma_5, \sigma_{\mu\nu} \} \ .
\end{align}
Using Fierz transformation this leads to%
\begin{subequations}
\begin{align}
 \Gamma^{3,XXY}_{\alpha\beta\gamma\delta} &= \Gamma^{4,XXY}_{\alpha\beta\gamma\delta} + \frac{1}{2} \Gamma^{9,XXY}_{\alpha\beta\gamma\delta} \ , \\
 \Gamma^{5,XXY}_{\alpha\beta\gamma\delta} &= 0 \ , \\
 \Gamma^{6,XXY}_{\alpha\beta\gamma\delta} &= \Gamma^{4,XXY}_{\alpha\beta\gamma\delta} - \Gamma^{7,XXY}_{\alpha\beta\gamma\delta}  \ ,
\end{align}
\end{subequations}
if $X\neq Y$. Therefore we have the freedom to choose%
\begin{align}
 \mathcal{F}^{3,j,k}_{\chodd}{(x_1,x_2,x_3)}&= 
 \mathcal{F}^{5,j,k}_{\chodd}{(x_1,x_2,x_3)}= 
 \mathcal{F}^{6,j,k}_{\chodd}{(x_1,x_2,x_3)}= 0 \ .
\end{align}
In the chiral-even sector the projection with $\gamma_{L/R}$ leads to similar constraints. The counterpart of eq.~\eqref{chiral_proj_cons_XXY} reads%
\begin{align}
  (\gamma_X \Gamma_A \gamma_X C)_{\gamma \beta} (\gamma_X \Gamma_B)_{\alpha \delta} &= 0 \ , \quad \text{if } \Gamma_A \in \{ \gamma_\mu, \gamma_\mu \gamma_5 \} \ .
\end{align}
With a Fierz transformation one obtains%
\begin{subequations}
\begin{align}
 \Gamma^{7,XXX}_{\alpha\beta\gamma\delta} &=  - \Gamma^{4,XXX}_{\alpha\beta\gamma\delta} + \frac{1}{2} \Gamma^{5,XXX}_{\alpha\beta\gamma\delta} + \Gamma^{6,XXX}_{\alpha\beta\gamma\delta}  \ , \\
  \Gamma^{8,XXX}_{\alpha\beta\gamma\delta} &= -\frac{1}{4}\Gamma^{5,XXX}_{\alpha\beta\gamma\delta} \ , \\
\Gamma^{9,XXX}_{\alpha\beta\gamma\delta} &= 0 \ .
\end{align}
\end{subequations}
Therefore, we can choose%
\begin{align}
  \mathcal{F}^{7,j,k}_{\cheven}{(x_1,x_2,x_3)}&=  
\mathcal{F}^{8,j,k}_{\cheven}{(x_1,x_2,x_3)}=  
\mathcal{F}^{9,j,k}_{\cheven}{(x_1,x_2,x_3)}= 0 \ .
\end{align}
The operators containing left-/right-handed quarks exclusively also have to be invariant under an exchange of the first and the last quark. Performing a Fierz transformation and using the identities given above we find%
\begin{align}
 \Gamma^{i,XXX}_{\gamma \beta \alpha \delta} & = \sum_{i^\prime=1}^6 \Gamma^{i^\prime,XXX}_{\alpha \beta \gamma \delta} c^{i^\prime i} \ .
\end{align}
The matrix $c$ is given by%
\begin{align}
 c&=\left( 
\begin{array}{*{6}Kc}
 \frac{1}{2} & 0 & -\frac{1}{2} & -\frac{3}{2} & 0 & -\frac{1}{2} \\
 0 & \frac{1}{2} & 0 & 0 & 6 & -\frac{1}{2} \\
 0 & 0 & 1 & 0 & 0 & 0 \\
 -\frac{1}{2} & 0 & -\frac{1}{2} & -\frac{1}{2} & 0 & -\frac{1}{2} \\
 0 & \frac{1}{8} & 0 & 0 & -\frac{1}{2} & \frac{1}{8} \\
 0 & 0 & 0 & 0 & 0 & 1
\end{array}
\right) .
\end{align}
By the use of this relation the symmetry property of the operator under exchange of the first and the last quark translates to the following constraints on the amplitudes:%
\begin{subequations} \label{symmetry_constraints_13}
\begin{align}
 &\text{\underline{$j=1,2$:}} & \mathcal{F}^{i,j,1}_{XX}{(x_1,x_2,x_3)} &= -\sum_{i^\prime=1}^6 c^{i i^\prime}\mathcal{F}^{i^\prime,j,3}_{XX} {(x_3,x_2,x_1)} \ , \\
 & & \mathcal{F}^{i,j,2}_{XX}{(x_1,x_2,x_3)} &= -\sum_{i^\prime=1}^6 c^{i i^\prime}\mathcal{F}^{i^\prime,j,2}_{XX} {(x_3,x_2,x_1)} \ , \\ 
 & & \mathcal{F}^{i,j,3}_{XX}{(x_1,x_2,x_3)} &= -\sum_{i^\prime=1}^6 c^{i i^\prime}\mathcal{F}^{i^\prime,j,1}_{XX} {(x_3,x_2,x_1)} \ ,\\[0.5cm]
 &\text{\underline{$j=3$:}} & \mathcal{F}^{i,3,1}_{XX}{(x_1,x_2,x_3)} &= -\sum_{i^\prime=1}^6 c^{i i^\prime}\mathcal{F}^{i^\prime,3,6}_{XX} {(x_3,x_2,x_1)} \ , \\
 & & \mathcal{F}^{i,3,2}_{XX}{(x_1,x_2,x_3)} &= -\sum_{i^\prime=1}^6 c^{i i^\prime}\mathcal{F}^{i^\prime,3,4}_{XX} {(x_3,x_2,x_1)} \ , \\
 & & \mathcal{F}^{i,3,3}_{XX}{(x_1,x_2,x_3)} &= -\sum_{i^\prime=1}^6 c^{i i^\prime}\mathcal{F}^{i^\prime,3,5}_{XX} {(x_3,x_2,x_1)} \ , \\
 & & \mathcal{F}^{i,3,4}_{XX}{(x_1,x_2,x_3)} &= -\sum_{i^\prime=1}^6 c^{i i^\prime}\mathcal{F}^{i^\prime,3,2}_{XX} {(x_3,x_2,x_1)} \ , \\
 & & \mathcal{F}^{i,3,5}_{XX}{(x_1,x_2,x_3)} &= -\sum_{i^\prime=1}^6 c^{i i^\prime}\mathcal{F}^{i^\prime,3,3}_{XX} {(x_3,x_2,x_1)} \ , \\
 & & \mathcal{F}^{i,3,6}_{XX}{(x_1,x_2,x_3)} &= -\sum_{i^\prime=1}^6 c^{i i^\prime}\mathcal{F}^{i^\prime,3,1}_{XX} {(x_3,x_2,x_1)} \ .
\end{align}
\end{subequations}
Using these equations one finds for the operator with $j = 3$ that one can eliminate all amplitudes apart from $\mathcal{F}^{i,3,1}_{XX}$, by using the following relations recursively:%
\begin{subequations}
\begin{align}
  \mathcal{F}^{i,3,2}_{XX}{(x_1,x_2,x_3)} &= \eta^C_{\Gamma^A,i} \sum_{i^\prime=1}^6  c^{i i^\prime} \mathcal{F}^{i^\prime,3,3}_{XX} {(x_3,x_1,x_2)} \ , \\
  \mathcal{F}^{i,3,3}_{XX}{(x_1,x_2,x_3)} &= \eta^C_{\Gamma^A,i} \sum_{i^\prime=1}^6  c^{i i^\prime} \mathcal{F}^{i^\prime,3,1}_{XX} {(x_3,x_1,x_2)} \ , \\
  \mathcal{F}^{i,3,4}_{XX}{(x_1,x_2,x_3)} &= -\eta^C_{\Gamma^A,i} \mathcal{F}^{i,3,1}_{XX} {(x_2,x_1,x_3)} \ , \\
  \mathcal{F}^{i,3,5}_{XX}{(x_1,x_2,x_3)} &= -\eta^C_{\Gamma^A,i} \mathcal{F}^{i,3,2}_{XX} {(x_2,x_1,x_3)} \ , \\
  \mathcal{F}^{i,3,6}_{XX}{(x_1,x_2,x_3)} &= -\eta^C_{\Gamma^A,i} \mathcal{F}^{i,3,3}_{XX} {(x_2,x_1,x_3)} \ ,
\end{align}
\end{subequations}
For the operators with $j =1,2$ we can eliminate%
\begin{subequations}
\begin{align}
 \mathcal{F}^{i,j,2}_{XX}{(x_1,x_2,x_3)}&= - \eta^C_{\Gamma^A,i}  \mathcal{F}^{i,j,1}_{XX}{(x_2,x_1,x_3)} \ , \\
 \mathcal{F}^{i,j,3}_{XX}{(x_1,x_2,x_3)}&= - \sum_{i^\prime=1}^6 c^{i i^\prime}  \mathcal{F}^{i^\prime,j,1}_{XX}{(x_3,x_2,x_1)} \ ,
\end{align}
\end{subequations}
and additionally%
\begin{subequations}
\begin{align}
\begin{split}
  \mathcal{F}^{1,j,1}_{XX}{(x_1,x_2,x_3)}&= \mathcal{F}^{3,j,1}_{XX}{(x_1,x_2,x_3)} + \mathcal{F}^{4,j,1}_{XX}{(x_1,x_2,x_3)} - 2 \mathcal{F}^{4,j,1}_{XX}{(x_1,x_3,x_2)} \\&\quad + \mathcal{F}^{6,j,1}_{XX}{(x_1,x_2,x_3)} \ ,
\end{split} \\
 \mathcal{F}^{2,j,1}_{XX}{(x_1,x_2,x_3)}&= -4 \mathcal{F}^{5,j,1}_{XX}{(x_1,x_2,x_3)} + 8 \mathcal{F}^{5,j,1}_{XX}{(x_1,x_3,x_2)} +  \mathcal{F}^{6,j,1}_{XX}{(x_1,x_2,x_3)} \ , \\
 \mathcal{F}^{3,j,1}_{XX}{(x_1,x_2,x_3)}&=-\mathcal{F}^{3,j,1}_{XX}{(x_1,x_3,x_2)} \ , \\
 \mathcal{F}^{6,j,1}_{XX}{(x_1,x_2,x_3)}&=-\mathcal{F}^{6,j,1}_{XX}{(x_1,x_3,x_2)} \ .
\end{align}
\end{subequations}
From the fact that the local operator at the origin, where $a_1=a_2=a_3=0$, is independent of the light-cone vector $n$ one can deduce constraints for the zeroth moments of the distribution amplitudes%
\begin{align} \label{zeroth_moments}
  \int [dx] \mathcal{F}^{i,j,k}_{XY}{(x_1,x_2,x_3)} &= 0 \ , \quad \text{for } i=1,3,4,6,7,9 \ .
\end{align}
\subsection{Elimination of linearly dependent structures} \label{sect_elim}
To avoid overparametrization we will now annihilate linearly dependent structures of those given in table~\ref{symmetry_properties_B}. Considering all possible three-quark operators and all baryons from the octet, one finds (for $j=1,2$) that only two out of the three structures $B^{j,1}_{\delta,a b c}$, $B^{j,2}_{\delta,a b c}$ and $B^{j,3}_{\delta,a b c}$ are linearly independent, since one has%
\begin{align} \label{lin_dep_j12}
 0&= B^{j,1}_{\delta,a b c} + B^{j,2}_{\delta,a b c} +  B^{j,3}_{\delta,a b c}  \ .
\end{align}
In the chiral-odd sector we can use this relation to replace $B^{j,3}_{\delta,a b c}=- B^{j,1}_{\delta,a b c} - B^{j,2}_{\delta,a b c} $, which is equivalent to the replacement%
\begin{subequations}
\begin{align}
 \mathcal{F}^{i,j,1}_{\chodd}{(x_1,x_2,x_3)} &\longrightarrow \tilde{\mathcal{F}}^{i,j,1}_{\chodd}{(x_1,x_2,x_3)} \equiv \mathcal{F}^{i,j,1}_{\chodd}{(x_1,x_2,x_3)} - \mathcal{F}^{i,j,3}_{\chodd}{(x_1,x_2,x_3)} \ , \\
 \mathcal{F}^{i,j,2}_{\chodd}{(x_1,x_2,x_3)} &\longrightarrow \tilde{\mathcal{F}}^{i,j,2}_{\chodd}{(x_1,x_2,x_3)} \equiv \mathcal{F}^{i,j,2}_{\chodd}{(x_1,x_2,x_3)} - \mathcal{F}^{i,j,3}_{\chodd}{(x_1,x_2,x_3)} \ , \\
 \mathcal{F}^{i,j,3}_{\chodd}{(x_1,x_2,x_3)} &\longrightarrow \tilde{\mathcal{F}}^{i,j,3}_{\chodd}{(x_1,x_2,x_3)} \equiv 0 \ .
\end{align}
\end{subequations}
Using eqs.~\eqref{symmetry_constraints_12_kj3_1} and~\eqref{symmetry_constraints_12_kj3_2} one finds that the new functions have the same symmetry properties as the old ones. Therefore we can choose%
\begin{align}
 \mathcal{F}^{i,j,3}_{\chodd}{(x_1,x_2,x_3)}&=0 \ , \quad j=1,2 \ ,
\end{align}
in accordance with symmetry properties and without loss of generality. In the chiral-even sector the situation is different since the amplitudes are already constrained by the symmetry under exchange of the first and the third quark. An elimination of one structure in favor of the two others would therefore not lead to a simplification. Instead one just obtains a reparametrization of the problem for which it would be hard to implement the symmetry properties under exchange of the first and the last quark.%
\section{Calculation at leading one-loop order} \label{sect_calculation}
In this section we describe the leading one-loop calculation. In section~\ref{sect_matching} we explain how we have matched to the standard DAs defined in ref.~\cite{Braun:2000kw}.
\subsection{Meson masses and the \texorpdfstring{$Z$}{Z}-factor}
\begin{figure}[t]
\centering%
\subfigure[\label{fdself:subfiga}]{
\includegraphics[scale=0.55]{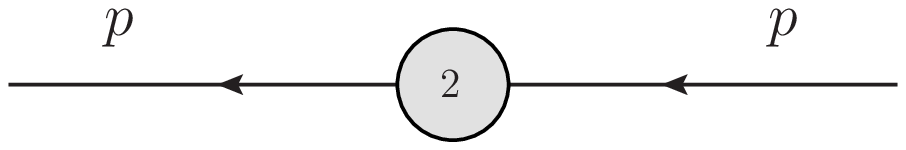}
}
\subfigure[\label{fdself:subfigb}]{
\includegraphics[scale=0.55]{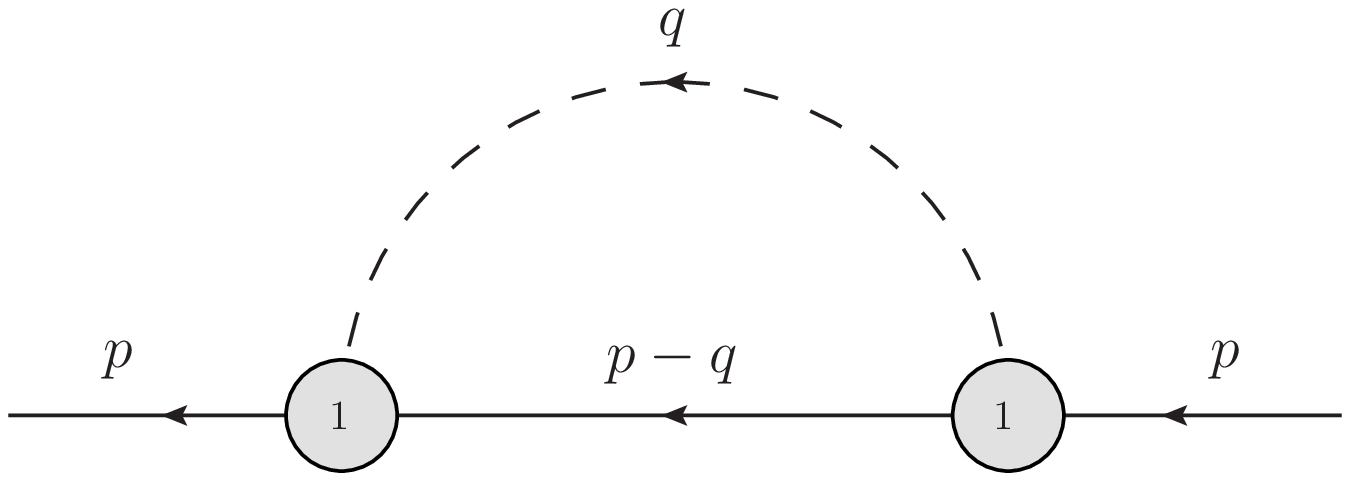}
}
\caption[]{\label{FD_selfenergy} Feynman diagrams needed for the calculation of the self-energy.}
\end{figure}

We work in the limit of exact isospin symmetry, where $m_u=m_d \equiv m_l$. Using the standard leading order meson Lagrangian (see e.g.~\cite{Gasser:1984gg,Scherer:2002tk}) one finds for the meson masses the standard Gell-Mann-Oakes-Renner relations%
\begin{subequations}
\begin{align}
 m_\pi^2  &= 2 B_0 m_l = m^2_{i=1,2,3} = 2 B_0 (\bar m_q - \delta m_l) \ , \\
 m_K^2    &= B_0 (m_l+m_s) = m^2_{i=4,\dots,7} = B_0 (2 \bar m_q + \delta m_l) \ , \\
 m_\eta^2 &=  \frac{B_0}{3} (2 m_l+ 4 m_s) = m^2_{i=8} = 2 B_0 ( \bar m_q + \delta m_l) \ ,
\end{align}
\end{subequations}
where%
\begin{subequations}
\begin{align}
\bar m_q & = \frac{1}{3} (2 m_l + m_s) \ , \\
\delta m_l & = \bar m_q - m_l \ ,
\end{align}
\end{subequations}
and $B_0$ is the LEC proportional to the quark condensate in the chiral limit. As additional ingredient we need the first order meson-baryon Lagrangian, which we take from ref.~\cite{Bruns:2011sh} (this version differs from refs.~\cite{Krause:1990xc,Scherer:2002tk} by a minus sign in the terms containing $D$ and $F$ in order to be consistent with the standard sign convention $g_A \approx D + F > 0$):%
\begin{align}
 \mathscr{L}_{MB}^{(1)} &= \tr{\bar B \gamma_\mu i D^\mu B} - m_0 \tr{\bar B B} + \frac{D}{2} \tr{\bar B \gamma_\mu \gamma_5 \{u^\mu,B\}} + \frac{F}{2} \tr{\bar B \gamma_\mu \gamma_5 [u^\mu,B]} \ .
\end{align}
For our calculation we need the baryon-meson-baryon vertex for an incoming baryon $B$, an outgoing baryon $B^\prime$ and an incoming meson (the $k$-th one in the Cartesian basis) with momentum $q$, which is given by%
\begin{align} \label{BMB_Vertex}
 \frac{-1}{2 F_0} \slashed{q} \gamma_5 \tr{\kappa_{B^\prime}^T (D \{\lambda^k,\kappa_B\}+F [\lambda^k,\kappa_B])} \ .
\end{align}
The self-energy to third chiral order is given by the sum of the irreducible diagrams shown in figure~\ref{FD_selfenergy} (where external legs are to be amputated) multiplied with an $i$. %
The contribution of diagram~\subref{fdself:subfigb} (in figure~\ref{FD_selfenergy}), which is relevant for the calculation of the $Z$-factor is given by%
\begin{align}
 i \times \text{\subref{fdself:subfigb}} &= 3 g_{B,\pi} f(m_\pi,m_0,\slashed{p}) + 4 g_{B,K} f(m_K,m_0,\slashed{p}) + g_{B,\eta} f(m_\eta,m_0,\slashed{p}) \ ,
\end{align}
where%
\begin{align}
f(m,m_0,\slashed{p}) &= \frac{-1}{4 F_0^2} \bigl( (p^2-m_0^2) \slashed{p} I^{(1)}_{11}{(m,m_0,\slashed{p})} + (\slashed{p}+m_0)\bigl( I_{01}{(m_0,\slashed{p})} - m^2 I_{11}{(m,m_0,\slashed{p})} \bigr) \bigr) \ .
\end{align}
The loop functions $I_{kl}$ and $I_{kl}^{(1)}$ are defined as in ref.~\cite{Wein:2011ix} and the coefficients are given by%
\begin{align}
 g_{N,\pi} &= (D+F)^2 \ , & g_{N,K} &= \frac{5}{6} D^2 - D F + \frac{3}{2} F^2 \ , & g_{N,\eta} &= \frac{1}{3} (D-3F)^2 \ , \nonumber \\
 g_{\Sigma,\pi} &= \frac{4}{9}(D^2 + 6 F^2) \ , & g_{\Sigma,K} &= D^2 + F^2 \ , & g_{\Sigma,\eta} &= \frac{4}{3} D^2 \ , \nonumber \\
 g_{\Xi,\pi} &=  (D-F)^2 \ , & g_{\Xi,K} &= \frac{5}{6} D^2 + D F + \frac{3}{2} F^2 \ , & g_{\Xi,\eta} &= \frac{1}{3} (D+3F)^2 \ , \nonumber \\
 g_{\Lambda,\pi} &= \frac{4}{3} D^2 \ , & g_{\Lambda,K} &= \frac{1}{3} (D^2 + 9 F^2) \ , & g_{\Lambda,\eta} &= \frac{4}{3} D^2 \ . \label{def_g_cefficients}
\end{align}
These constants fulfill the constraints that the sums%
\begin{subequations}
\begin{align}
3 g_{B,\pi} + 4 g_{B,K} + g_{B,\eta} &= \frac{4}{3} (5 D^2+9 F^2) \ , \\
2 g_{N,M} + 3 g_{\Sigma,M} + 2 g_{\Xi,M} + g_{\Lambda,M}  &= \frac{4}{3} (5 D^2+9 F^2) \ ,
\end{align}
\end{subequations}
are independent of the baryon/meson species. This yields similar baryon masses and $Z$-factors along the line of equal quark masses and is a consequence of $\operatorname{SU}(3)_f$ symmetry. For a detailed study of baryon masses under symmetry breaking see~\cite{Bruns:2012eh}. The square root of the $Z$-factor needed in our calculation is given by%
\begin{align}
 \sqrt{Z_B} &\mathrel{\dot =} 1+\frac{1}{2}\Sigma^\prime_B \ ,
\end{align}
where the prime indicates taking a derivative with respect to $\slashed p$ and substituting $\slashed{p}\rightarrow m_B$, while%
\begin{align}
\begin{split}
\Sigma^\prime_B(\bar m_q, \delta m_l) &= 3 g_{B,\pi} f^\prime(m_\pi,m_0,m_B) + 4 g_{B,K} f^\prime(m_K,m_0,m_B) + g_{B,\eta} f^\prime(m_\eta,m_0,m_B) \\
&\equiv \Sigma^{\prime\star}(\bar m_q) + \Delta \Sigma^\prime_B(\bar m_q, \delta m_l) \ ,
\end{split}
\end{align}
with%
\begin{subequations}
\begin{align}
\begin{split}
 \Sigma^{\prime\star}(\bar m_q) &= \Sigma^\prime_B(\bar m_q, 0) = (3 g_{B,\pi}+4 g_{B,K}+g_{B,\eta}) f^\prime(m_m^\star,m_0,m_B) \\
 &= \frac{4}{3} (5 D^2+9 F^2) f^\prime(m_m^\star,m_0,m_B) \mathrel{\dot =}\frac{4}{3} (5 D^2+9 F^2)  f^\prime(m_m^\star,m_b^\star,m_b^\star) \ ,
\end{split} \raisetag{1cm} \\
\begin{split}
 \Delta\Sigma^{\prime}_B(\bar m_q, \delta m_l) &= \Sigma^\prime_B(\bar m_q, \delta m_l) -\Sigma^\prime_B(\bar m_q, 0) \\
&=  3 g_{B,\pi} f^\prime(m_\pi,m_0,m_B) + 4 g_{B,K} f^\prime(m_K,m_0,m_B) + g_{B,\eta} f^\prime(m_\eta,m_0,m_B)\\
&\quad - \frac{4}{3} (5 D^2+9 F^2) f^\prime(m_m^\star,m_0,m_B) \\
&\mathrel{\dot =} 3 g_{B,\pi} f^\prime(m_\pi,m_b^\star,m_b^\star) + 4 g_{B,K} f^\prime(m_K,m_b^\star,m_b^\star) + g_{B,\eta} f^\prime(m_\eta,m_b^\star,m_b^\star)\\
&\quad - \frac{4}{3} (5 D^2+9 F^2) f^\prime(m_m^\star,m_b^\star,m_b^\star) \ ,
\end{split} \raisetag{0.4cm}
\end{align}
\end{subequations}
where $m_{m/b}^\star = m_{m/b}^\star (\bar m_q)$ is the meson/baryon mass along the symmetric line ($\delta m_l = 0$). The dotted equal sign $\mathrel{\dot =}$ means equal up to terms which are of higher order than our level of accuracy (which is second order in chiral power counting). For explicit results see appendix~\ref{sect_gfunctions}.%
\subsection{Baryon-to-vacuum matrix elements of three-quark operators}
In this section we describe the actual loop calculation. From a simple power counting argument one finds that at leading one-loop level the only contributing graphs are the ones shown in figure~\ref{FDDA}. One easily observes that the second order operator insertions only occur without additional mesons. Therefore we only have to compute the vertices where a single baryon couples to the operator. Contributions with additional mesons only occur for the leading order operator insertion ($j=1$). For the BChPT calculation mainly the structure $B^{j,k,XYZ}_{\delta,abc}$ is relevant. Graph \subref{fdda:subfigd} of figure~\ref{FDDA} is an exception because the extra $\gamma_5$ from the baryon-meson-baryon vertex has to be canceled with a $\gamma_5$ from the Dirac structure of the operator. The calculation gets simplified considerably if one uses the fact that (by construction) the $B^{j,k,XYZ}_{\delta,abc}$ with $k\neq 1$ can be obtained from the case $k=1$ by a permutation of indices:%
\begin{figure}[t]
\centering%
\subfigure[\label{fdda:subfiga}]{
\includegraphics[scale=0.55]{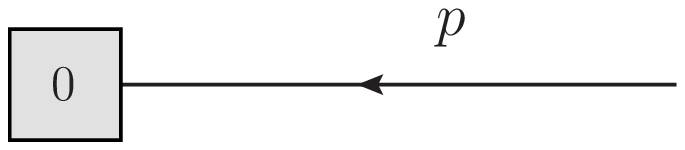}
}
\subfigure[\label{fdda:subfigb}]{
\includegraphics[scale=0.55]{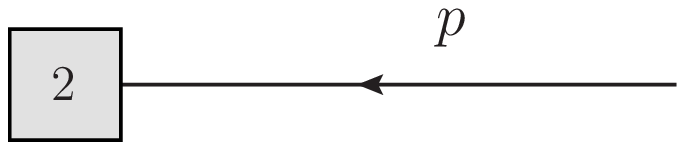}
}
\subfigure[\label{fdda:subfigc}]{
\includegraphics[scale=0.55]{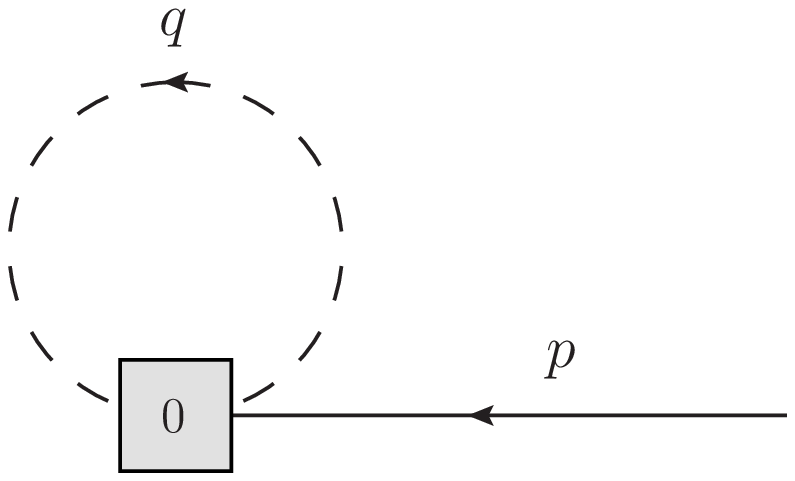}
}
\subfigure[\label{fdda:subfigd}]{
\includegraphics[scale=0.55]{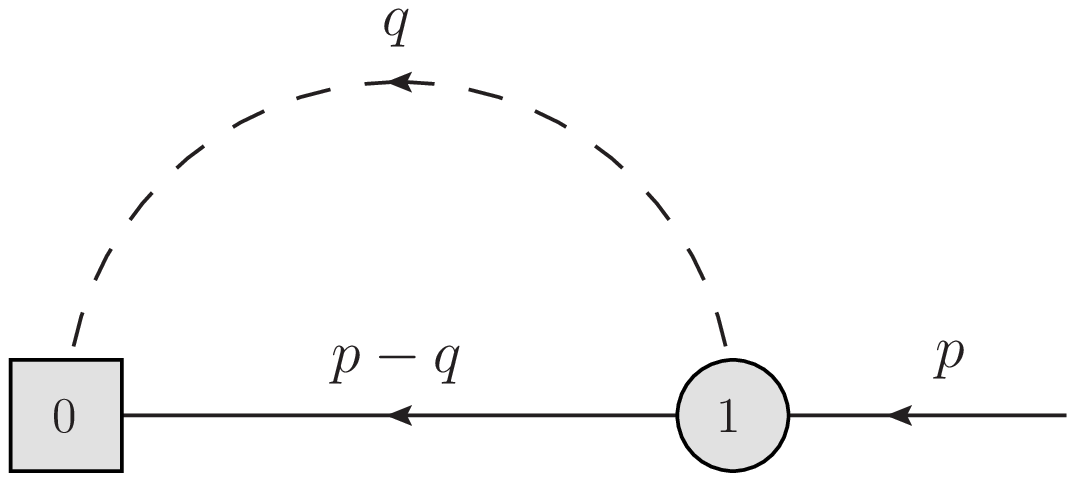}
}
\caption[]{\label{FDDA} Feynman diagrams needed for the calculation of the baryon-to-vacuum matrix elements. The squares depict the operator insertions given in eqs.~\eqref{operator_insertions}, the circle stands for the vertex from the meson-baryon Lagrangian given in eq.~\eqref{BMB_Vertex} and the dashed/solid lines represent mesons/baryons. Diagram \subref{fdda:subfiga} has to be multiplied with $\sqrt{Z}$. However, one knows that at higher orders all of the diagrams will receive a $\sqrt{Z}$ contribution, which can be used as an argument in favor of the factorized version of our results (see eq.~\eqref{DAs_chpt_Result_factorized} in section~\ref{sect_result}).}
\end{figure}%
\begin{align}
 B^{j,2,XYZ}_{\delta,abc}&=B^{j,1,YZX}_{\delta,bca} \ , &
 B^{j,3,XYZ}_{\delta,abc}&=B^{j,1,ZXY}_{\delta,cab} \ , &
 B^{j,4,XYZ}_{\delta,abc}&=-B^{j,1,YXZ}_{\delta,bac} \ , \nonumber \\*
 B^{j,5,XYZ}_{\delta,abc}&=-B^{j,1,XZY}_{\delta,acb} \ , &
 B^{j,6,XYZ}_{\delta,abc}&=-B^{j,1,ZYX}_{\delta,cba} \ , 
\end{align}
which means that we actually only have to calculate the case $k=1$. Defining%
\begin{align}
 (-1)_X &\equiv \begin{cases}
            +1 \ , & \text{for } X=R \\ -1 \ , & \text{for } X=L
           \end{cases} \ ,
\end{align}
we can write down the relevant operator insertions in a quite economic way:%
\begin{subequations} \label{operator_insertions}
\begin{align}
B^{1,1,XYZ}_{\delta,abc}(z) \bigg|_{B_\epsilon (p)}&= (\kappa_B)^{a a^\prime} \varepsilon^{a^\prime b c} e^{-i p \cdot z} \delta_{\delta \epsilon} \ , \\
B^{1,1,XYZ}_{\delta,abc}(z) \bigg|_{B_\epsilon (p-q) \phi^k(q)}&= \frac{i}{2 F_0} \Bigl[ \begin{aligned}[t] & (-1)_X (\lambda^k \kappa_B)^{a a^\prime} \delta^{bb^\prime} \delta^{cc^\prime} +  (-1)_Y ( \kappa_B)^{a a^\prime} (\lambda^k)^{bb^\prime} \delta^{cc^\prime} \\ &  +  (-1)_Z (\kappa_B)^{a a^\prime} \delta^{bb^\prime} (\lambda^k )^{cc^\prime}  \Bigr] \varepsilon^{a^\prime b^\prime c^\prime} e^{-i p \cdot z} \delta_{\delta \epsilon} \ , \end{aligned}
\end{align}
\begin{align}
\begin{split}
\MoveEqLeft B^{1,1,XYZ}_{\delta,abc}(z) \bigg|_{B_\epsilon (p-q_1-q_2) \phi^k(q_1) \phi^l(q_2)}=\\
&= \frac{-1}{16 F_0^2} \Bigl[ \begin{aligned}[t] & (\{\lambda^k,\lambda^l\} \kappa_B)^{a a^\prime} \delta^{bb^\prime} \delta^{cc^\prime} +  ( \kappa_B)^{a a^\prime} (\{\lambda^k,\lambda^l\})^{bb^\prime} \delta^{cc^\prime}  +  (\kappa_B)^{a a^\prime} \delta^{bb^\prime} (\{\lambda^k,\lambda^l\})^{cc^\prime} \\
&+ 2 (-1)_X (-1)_Y (\lambda^k \kappa_B)^{a a^\prime} (\lambda^l)^{bb^\prime} \delta^{cc^\prime}
+ 2 (-1)_X (-1)_Z (\lambda^k \kappa_B)^{a a^\prime} \delta^{bb^\prime} (\lambda^l)^{cc^\prime} \\
&+ 2 (-1)_Y (-1)_Z (\kappa_B)^{a a^\prime} (\lambda^k )^{bb^\prime} (\lambda^l)^{cc^\prime}
+ 2 (-1)_X (-1)_Y (\lambda^l \kappa_B)^{a a^\prime} (\lambda^k)^{bb^\prime} \delta^{cc^\prime}\\
&+ 2 (-1)_X (-1)_Z (\lambda^l \kappa_B)^{a a^\prime} \delta^{bb^\prime} (\lambda^k)^{cc^\prime} 
+ 2 (-1)_Y (-1)_Z (\kappa_B)^{a a^\prime} (\lambda^l )^{bb^\prime} (\lambda^k)^{cc^\prime} 
 \Bigr] \\
& \times  \varepsilon^{a^\prime b^\prime c^\prime} e^{-i p \cdot z} \delta_{\delta \epsilon} \ . \end{aligned} 
\end{split}
\end{align}
The second order tree-level operator insertions read%
\begin{align}
 B^{2,1,XYZ}_{\delta,abc}(z) \bigg|_{B_\epsilon (p)}&= 4 B_0 m_0^{-2} \tr{\mathcal{M}}(\kappa_B)^{a a^\prime} \varepsilon^{a^\prime b c} e^{-i p \cdot z} \delta_{\delta \epsilon} \ , \\ 
 B^{3,1,XYZ}_{\delta,abc}(z) \bigg|_{B_\epsilon (p)}&= 4 B_0 m_0^{-2} (\kappa_B)^{a a^\prime} \tilde{\mathcal{M}}^{b b^\prime} \varepsilon^{a^\prime b^\prime c} e^{-i p \cdot z} \delta_{\delta \epsilon} \ ,
\end{align}
\end{subequations}
where $\tilde{\mathcal{M}}= \mathcal{M}-\smash{\tr{\mathcal{M}}}/3$. After performing the loop calculation one finds that the results can be expressed as%
\begin{subequations}
\begin{align}
\begin{split}
\MoveEqLeft \langle 0 | \mathcal O^{abc}_{RR,\alpha \beta \gamma} {(a_1,a_2,a_3)} + \mathcal O^{abc}_{LL,\alpha \beta \gamma}{(a_1,a_2,a_3)}| B(p,s) \rangle= \\
&= \int [dx] e^{- i \; n \cdot p \sum_k x_k a_k} \sum_i \Gamma^{i,\cheven}_{\alpha\beta\gamma\delta} u^B_\delta (p,s) h_{B,\cheven}^{i,abc}{(x_1,x_2,x_3)} \ ,
\end{split} \\
\begin{split}
\MoveEqLeft\langle 0 | \mathcal O^{abc}_{RL,\alpha \beta \gamma} {(a_1,a_2,a_3)} + \mathcal O^{abc}_{LR,\alpha \beta \gamma}{(a_1,a_2,a_3)}| B(p,s) \rangle =\\
&= \int [dx] e^{- i \; n \cdot p  \sum_k x_k a_k} \sum_i \Gamma^{i,\chodd}_{\alpha\beta\gamma\delta} u^B_\delta (p,s) h_{B,\chodd}^{i,abc}{(x_1,x_2,x_3)} \ ,
\end{split}
\end{align}
\end{subequations}
where $u^B_\delta (p,s)$ is the baryon spinor,%
\begin{subequations} \label{gamma_structure_odd_even}
 \begin{align}
  \Gamma^{i,\cheven}_{\alpha\beta\gamma\delta}  &= \Gamma^{i,RRR}_{\alpha\beta\gamma\delta}-\Gamma^{i,LLL}_{\alpha\beta\gamma\delta} \ , \\
  \Gamma^{i,\chodd}_{\alpha\beta\gamma\delta}   &= \Gamma^{i,LLR}_{\alpha\beta\gamma\delta}-\Gamma^{i,RRL}_{\alpha\beta\gamma\delta} \ ,
 \end{align}
\end{subequations}
and%
\begin{subequations} \label{def_h_functions}
 \begin{align}
  h^{i,abc}_{B,\cheven}{(x_1,x_2,x_3)}  &= \sum_{j,k}  c^{j,k,abc}_{B,RRR} \mathcal F^{i,j,k}_\cheven {(x_1,x_2,x_3)}  \ , \\
  h^{i,abc}_{B,\chodd} {(x_1,x_2,x_3)}  &= \sum_{j,k}  c^{j,k,abc}_{B,LLR} \mathcal F^{i,j,k}_\chodd {(x_1,x_2,x_3)} \ .
 \end{align}
\end{subequations}
The coefficients $c^{j,k,abc}_{B,XYZ}$ inherit the property that the ones with $k\neq 1$ can be obtained from the case $k=1$ by a permutation of indices:%
\begin{align}
 c^{j,2,abc}_{B,XYZ}&= c^{j,1,bca}_{B,YZX} \ , &
 c^{j,3,abc}_{B,XYZ}&= c^{j,1,cab}_{B,ZXY} \ , &
 c^{j,4,abc}_{B,XYZ}&=- c^{j,1,bac}_{B,YXZ} \ , \nonumber \\*
 c^{j,5,abc}_{B,XYZ}&=- c^{j,1,acb}_{B,XZY} \ , &
 c^{j,6,abc}_{B,XYZ}&=- c^{j,1,cba}_{B,ZYX} \ .
\end{align}
For those with $k=1$ we find%
\begin{subequations}
\begin{align}
  c^{1,1,abc}_{B,XYZ} &= c^{1,1,abc}_{B,XYZ}\bigg|_{(a)} +c^{1,1,abc}_{B,XYZ}\bigg|_{(c)} + c^{1,1,abc}_{B,XYZ}\bigg|_{(d)} \ , \\
  c^{1,1,abc}_{B,XYZ}\bigg|_{(a)} &= \sqrt{Z_B} (\kappa_B)^{aa^\prime} \varepsilon^{a^\prime b c} \ , \\
  c^{1,1,abc}_{B,XYZ}\bigg|_{(c)} &= \frac{-1}{8 F_0^2} \sum_k  
\Bigl[ \begin{aligned}[t] & (\lambda^k \lambda^k \kappa_B)^{a a^\prime} \delta^{bb^\prime} \delta^{cc^\prime} +  ( \kappa_B)^{a a^\prime} (\lambda^k \lambda^k)^{bb^\prime} \delta^{cc^\prime}  +  (\kappa_B)^{a a^\prime} \delta^{bb^\prime} (\lambda^k \lambda^k)^{cc^\prime} \\
&+ 2 (-1)_X (-1)_Y (\lambda^k \kappa_B)^{a a^\prime} (\lambda^k)^{bb^\prime} \delta^{cc^\prime} \\
&+ 2 (-1)_X (-1)_Z (\lambda^k \kappa_B)^{a a^\prime} \delta^{bb^\prime} (\lambda^k)^{cc^\prime} \\
&+ 2 (-1)_Y (-1)_Z (\kappa_B)^{a a^\prime} (\lambda^k )^{bb^\prime} (\lambda^k)^{cc^\prime}
 \Bigr] \varepsilon^{a^\prime b^\prime c^\prime} I_{10}{(m_l)}  \ , \end{aligned} \\
\begin{split}
  c^{1,1,abc}_{B,XYZ}\bigg|_{(d)} &= \frac{-1}{4 F_0^2} \sum_{k,{\tilde{B}}}
\Bigl[ \begin{aligned}[t] & (-1)_X (\lambda^k \kappa_{\tilde{B}})^{a a^\prime} \delta^{bb^\prime} \delta^{cc^\prime} +(-1)_Y ( \kappa_{\tilde{B}})^{a a^\prime} (\lambda^k)^{bb^\prime} \delta^{cc^\prime} \\
& +(-1)_Z ( \kappa_{\tilde{B}})^{a a^\prime} \delta^{bb^\prime} (\lambda^k)^{cc^\prime} \Bigr] \tr{\kappa_{\tilde{B}}^T(D\{\lambda^k,\kappa_B\}+F[\lambda^k,\kappa_B])} \end{aligned} \\
& \ \times\Bigl( I_{10}{(m_k)} + (m_B^2-m_0^2) I_{11}{(m_k,m_0,m_B)} - m_B(m_B+m_0) I^{(1)}_{11}{(m_k,m_0,m_B)} \Bigr) \ .
\end{split}
\end{align}
In the contribution from graph \subref{fdda:subfigd} commuting $\gamma_5$ from the vertex with the Dirac structure in the operator yields $\eta^5_{\Gamma^B,i}(-1)^{d^m_i}=-1$ (compare table~\ref{symmetry_properties_Gamma}). In operators of type $\mathcal O_{XR}$ the $\gamma_5$ has no effect owing to $\gamma_R \gamma_5=\gamma_R$. The relative sign in the vertex in operators of type $\mathcal O_{\bar XL}$ is compensated by $\gamma_L \gamma_5 = - \gamma_L$. Therefore the result only contains structures of the form given in eq.~\eqref{gamma_structure_odd_even}. This is no coincidence but has to happen in order to obtain a result that behaves correctly under parity transformation. For the second order tree-level contributions we find%
\begin{align}
 c^{2,1,abc}_{B,XYZ} &=  4 B_0 m_0^{-2} \tr{\mathcal{M}} (\kappa_B)^{a a^\prime} \varepsilon^{a^\prime b c}  \ , \\ 
 c^{3,1,abc}_{B,XYZ} &=  4 B_0 m_0^{-2} (\kappa_B)^{a a^\prime} \mathcal{\tilde M}^{bb^\prime} \varepsilon^{a^\prime b^\prime c}  \ .
\end{align}
\end{subequations}
Using eq.~\eqref{3q_op_decompRL} the matrix element of the complete three-quark operator reads%
\begin{align} \label{result_not_projected}
\begin{split}
 \MoveEqLeft[1]\langle 0 | q^a_\alpha (a_1 n) q^b_\beta (a_2 n) q^c_\gamma (a_3 n) | B(p,s) \rangle= \\
&= \int [dx] e^{- i \; n \cdot p  \sum_k x_k a_k} \sum_i 
 \Bigl(  \begin{aligned}[t] & \Gamma^{i,\cheven}_{\alpha\beta\gamma\delta}  h_{B,\cheven}^{i,abc}{(x_1,x_2,x_3)} + \Gamma^{i,\chodd}_{\alpha\beta\gamma\delta}  h_{B,\chodd}^{i,abc}{(x_1,x_2,x_3)}\\
& + \Gamma^{i,\chodd}_{\gamma\alpha\beta\delta}  h_{B,\chodd}^{i,cab}{(x_3,x_1,x_2)} + \Gamma^{i,\chodd}_{\beta\gamma\alpha\delta}  h_{B,\chodd}^{i,bca}{(x_2,x_3,x_1)} \Bigr)  u^B_\delta (p,s) \ . \end{aligned} 
\end{split}
\end{align}
\subsection{Projection onto standard DAs} \label{sect_matching}
In this section we relate our parametrization of the baryon-to-vacuum matrix element, which was guided by the behaviour under chiral rotations, to the general decomposition given in ref.~\cite{Braun:2000kw}, which is more convenient for daily use. To do so we have contracted both our result (eq.~\eqref{result_not_projected}) and formula (2.3) of ref.~\cite{Braun:2000kw} with Dirac structures of the form $\Gamma^A_{\alpha\beta}\otimes \Gamma^B_{\gamma^\prime \gamma}$. It is sufficient to use structures where Lorentz indices are either contracted between $\Gamma^A$ and $\Gamma^B$ or with the light-cone vector $n$ or the momentum $p$. Afterwards we have used the identity $\slashed{p} u^B(p)=m_B u^B(p)$ and have matched the prefactors of the remaining Dirac structures ($\gamma_5$ and $\slashed{n} \gamma_5$). Using twist-projection, we obtain the results for the distribution amplitudes $S_i^B$, $P_i^B$, $A_i^B$, $V_i^B$ and $T_i^B$ which are independent of the scalar product $n \cdot p$, due to the scaling property described in section~\ref{sect_sym_3qopp} (For the details of the twist-projection we refer to \cite{Braun:2000kw}). We have collected these lengthy matching relations in appendix~\ref{app_projection_standard_DAs}. The amplitudes have the following symmetry properties under exchange of the first and the second variable%
 \begin{align} 
  S_i^B{(x_1,x_2,x_3)} &=  - (-1)_B S_i^B{(x_2,x_1,x_3)} \ , \nonumber \\
  P_i^B{(x_1,x_2,x_3)} &=  - (-1)_B P_i^B{(x_2,x_1,x_3)} \ , \nonumber \\
  A_i^B{(x_1,x_2,x_3)} &=  - (-1)_B A_i^B{(x_2,x_1,x_3)} \ , \nonumber \\
  V_i^B{(x_1,x_2,x_3)} &=  + (-1)_B V_i^B{(x_2,x_1,x_3)} \ , \nonumber \\*
  T_i^B{(x_1,x_2,x_3)} &=  + (-1)_B T_i^B{(x_2,x_1,x_3)} \ , \label{symmetry_standard_DAs_12}
 \end{align}
where we use%
\begin{align}
 (-1)_B &\equiv \begin{cases} +1 \ , & \text{for } B \neq \Lambda \\ -1 \ , & \text{for } B = \Lambda \end{cases} \ ,
\end{align}
for brevity. To obtain these nice symmetry properties one has to choose the flavor content in the operator as $p \mathrel{ \hat = } uud$, $n \mathrel{ \hat = } ddu$, $\Sigma^+ \mathrel{ \hat = } uus$, $\Sigma^0  \mathrel{ \hat = } uds$, $\Sigma^- \mathrel{ \hat = } dds$, $\Xi^0 \mathrel{ \hat = } ssu$, $\Xi^- \mathrel{ \hat = } ssd$, $\Lambda \mathrel{ \hat = } uds$, where the order of the flavors is relevant. The different sign for the $\Lambda$ originates from the antisymmetry of the isospin singlet state.%
\section{Results} \label{sect_result}
In this section we present our results and provide a definition for DAs that do not mix under chiral extrapolation. In section~\ref{sect_example_of_application} we work out an explicit parametrization of baryon octet DAs, where we follow the approach presented in refs.~\cite{Anikin:2013aka,Anikin:2013yoa,Braun:2008ia}.
\subsection{General strategy and choice of distribution amplitudes}
We will split up every distribution amplitude in the following way:%
\begin{subequations}
\begin{align}
\begin{split}
\text{DA}(\bar m_q, \delta m_l) &= \text{DA}(\bar m_q, 0) + \bigl(\text{DA}(\bar m_q, \delta m_l) - \text{DA}(\bar m_q, 0)\bigr) \\
&\equiv \text{DA}^\star (\bar m_q) + \Delta \text{DA} (\bar m_q, \delta m_l) \ ,
\end{split} \\
\begin{split}
\text{DA}^\star(\bar m_q) &= \text{DA}^\star(0) + \bigl(\text{DA}^\star(\bar m_q) - \text{DA}^\star(0)\bigr) \\
&\equiv \text{DA}^\circ + \Delta \text{DA}^\star(\bar m_q) \ ,
\end{split} 
\end{align}
\end{subequations}
where the main idea is to use the second formula to parametrize everything in terms of the DAs at the symmetric point, which are measurable on the lattice as opposed to the amplitudes in the chiral limit. Lattice simulations where the mean quark mass is fixed at its physical value while $\delta m_l$ is varied are already available for hadron masses and some form factors~\cite{Bietenholz:2011qq,Gockeler:2011ze,Cooke:2013qqa}. Corresponding simulations for the baryon octet DAs treated in this work are in progress. This strategy has the additional advantage that one gets rid of the parameters that describe the behaviour under variation of the mean quark mass. For the presentation of the results it turns out to be convenient to write down the second order tree-level and the loop contribution separately. We define for all baryons%
\begin{align}
 \Delta \text{DA} = \Delta \text{DA}^{\text{loop}} + \delta m \ \Delta \text{DA}^{\text{tree}} \ ,
\end{align}
where
\begin{align}
 \delta m&=\frac{4 B_0 \delta m_l}{{m_b^\star}^2} \ .
\end{align}
Then we use the fact that we can rewrite $\Delta \text{DA}$ in terms of $m_b^\star$ and $\text{DA}^\star$ using the corresponding expansions in $\bar m_q$. For a specific set of DAs, which do not mix under chiral extrapolation (see below), this allows us to rewrite the loop contribution as the DA along the symmetric line multiplied with a loop function $f$ such that the results have the form%
\begin{align}
\begin{split}
\text{DA}(\bar m_q, \delta m_l) &= \text{DA}^\star (\bar m_q) (1+f) + \delta m \ \Delta \text{DA}^{\text{tree}} \ .
\end{split}
\end{align}
By virtue of $\operatorname{SU}(3)_f$ symmetry we find the following relations between DAs along the line of symmetric quark masses $m_u=m_d=m_s$:%
\begin{subequations} \label{su3constraints}
 \begin{align}
\begin{split}
  2 T_{1/6}^{B\star} { (x_1,x_2,x_3)} &= (-1)_B \bigl[V_{1/6}^{B\star}-A_{1/6}^{B\star}\bigr]{ (x_1,x_3,x_2)} \\&\quad+ \bigl[V_{1/6}^{B\star}-A_{1/6}^{B\star}\bigr]{ (x_2,x_3,x_1)} \ ,
\end{split} \\*
\begin{split}
  \bigl[T_{3/4}^{B\star} + T_{7/8}^{B\star} + S_{1/2}^{B\star} - P_{1/2}^{B\star}\bigr]{ (x_1,x_2,x_3)} &= \bigl[V_{2/5}^{B\star}-A_{2/5}^{B\star}\bigr]{ (x_2,x_3,x_1)} \\&\quad+ \bigl[V_{3/4}^{B\star}-A_{3/4}^{B\star}\bigr]{ (x_3,x_1,x_2)} \ ,
\end{split} \\*
\begin{split}
 2 T_{2/5}^{B\star} { (x_1,x_2,x_3)} &= \bigl[T_{3/4}^{B\star} - T_{7/8}^{B\star} + S_{1/2}^{B\star} + P_{1/2}^{B\star}\bigr]{ (x_3,x_1,x_2)}\\&\quad + \bigl[T_{3/4}^{B\star} - T_{7/8}^{B\star} + S_{1/2}^{B\star} + P_{1/2}^{B\star}\bigr]{ (x_3,x_2,x_1)} \ .
\end{split} 
\end{align}
\end{subequations}
Note that we do not impose these relations. They are automatically fulfilled by our calculation (loop contributions included). For the nucleons these relations are fulfilled exactly also for $\delta m_l \neq 0$ owing to isospin symmetry (again this is also true for the loop contributions), which was already shown in ref.~\cite{Braun:2000kw}. If we were only interested in the $\operatorname{SU}(3)_f$ symmetric case (or in nucleons only), it would therefore be enough to define the independent amplitudes as%
\begin{subequations}\label{original_DAs_definition}
\begin{align}
\Phi_{3/6}^B(x_1,x_2,x_3)&=\bigl[ V_{1/6}^B-A_{1/6}^B \bigr](x_1,x_2,x_3) \ , \\
   \Phi_{4/5}^B(x_1,x_2,x_3)&=\bigl[ V_{2/5}^B-A_{2/5}^B \bigr](x_1,x_2,x_3) \ , \\
   \Psi_{4/5}^B(x_1,x_2,x_3)&=\bigl[ V_{3/4}^B-A_{3/4}^B \bigr](x_1,x_2,x_3) \ , \\
   \Xi_{4/5}^B(x_1,x_2,x_3)&=\bigl[ T_{3/4}^B-T_{7/8}^B+S_{1/2}^B+P_{1/2}^B \bigr](x_1,x_2,x_3) \ ,
\end{align}
\end{subequations}
where the $\Phi_i^B$ and $\Psi_i^B$ describe the coupling to chiral-odd operators, while the $\Xi_i^B$ describe the chiral-even sector. The subscript indicates the twist. As it turns out the amplitudes $\Phi_i^B$, $\Psi_i^B$ and $\Xi_i^B$ are not yet the optimal choice for a description of the complete baryon octet, since they mix under chiral extrapolation. Additionally one finds that it is very convenient to use differing definitions for the $\Lambda$, which we choose in such a way that the DAs of the $\Lambda$ coincide with the DAs of the other octet baryons in the limit of equal quark masses. Therefore we define%
\begin{subequations} \label{definition_superior_DAs}
 \begin{align}
   \Phi_{\pm, 3/6}^B(x_1,x_2,x_3)&=\frac{c^\pm_B}{2} \bigl(\bigl[ V_{1/6}^B-A_{1/6}^B \bigr](x_1,x_2,x_3) \pm \bigl[ V_{1/6}^B-A_{1/6}^B \bigr](x_3,x_2,x_1) \bigr) \ , \\
   \Phi_{\pm, 4/5}^B(x_1,x_2,x_3)&=c^\pm_B \bigl(\bigl[ V_{2/5}^B-A_{2/5}^B \bigr](x_1,x_2,x_3) \pm (-1)_B \bigl[ V_{3/4}^B-A_{3/4}^B \bigr](x_2,x_3,x_1)\bigr) \ , \\
   \Xi_{\pm, 4/5}^B(x_1,x_2,x_3)&=3(-1)_B c^\pm_B \bigl( \begin{aligned}[t] & \bigl[ T_{3/4}^B-T_{7/8}^B+S_{1/2}^B+P_{1/2}^B \bigr](x_1,x_2,x_3) \\& \pm \bigl[ T_{3/4}^B-T_{7/8}^B+S_{1/2}^B+P_{1/2}^B \bigr](x_1,x_3,x_2) \bigr) \ , \end{aligned} 
\end{align}
\end{subequations}%
where%
\begin{align}
c^+_B & = \begin{cases} 1 \ , & \text{for } B\neq\Lambda \\ \sqrt{\frac{2}{3}} \ , & \text{for } B=\Lambda\end{cases} \ , & c^-_B & = \begin{cases} 1 \ , & \text{for } B\neq\Lambda \\ -\sqrt{6} \ , & \text{for } B=\Lambda \end{cases} \ .
\end{align}
Being interested in $\operatorname{SU}(3)_f$ violation one can not use the constraints given in eq.~\eqref{su3constraints} and therefore one needs six additional DAs. Our choice are (up to differing prefactors for the $\Lambda$ and exchange of variables) the left-hand sides in eq.~\eqref{su3constraints} since they coincide with the DAs in eq.~\eqref{definition_superior_DAs} in the $\operatorname{SU(3)}_f$ symmetric limit. We define%
\begin{subequations}  \label{definition_superior_DAs_second_part}
 \begin{align}
   \Pi_{3/6}^B(x_1,x_2,x_3)&=c_B^- (-1)_B \; T_{1/6}^B (x_1,x_3,x_2) \ , \\
   \Pi_{4/5}^B(x_1,x_2,x_3)&=c_B^- \bigl[T_{3/4}^B + T_{7/8}^B + S_{1/2}^B - P_{1/2}^B\bigr](x_3,x_1,x_2) \ , \\
  \Upsilon_{4/5}^B(x_1,x_2,x_3)&= 6 c_B^- \;  T_{2/5}^B(x_3,x_2,x_1) \ ,  
\end{align}%
\end{subequations}%
where the $\Pi_i$ describe the chiral-odd sector, while the $\Upsilon_i$ describe the chiral-even part. For each octet baryon the standard DAs can be decomposed into the amplitudes defined in eqs.~\eqref{definition_superior_DAs} and~\eqref{definition_superior_DAs_second_part} (see appendix~\ref{app_handbook_of_DAs}). We find that the DAs for different nucleons, $\Sigma$'s and $\Xi$'s are related to each other exactly by isospin symmetry. Therefore we define%
\begin{subequations} \label{baryon_signs}
 \begin{align}
  \text{DA}^N &\equiv \text{DA}^p = - \text{DA}^n \ , \\
  \text{DA}^\Sigma &\equiv \text{DA}^{\Sigma^-} = - \text{DA}^{\Sigma^+} = \sqrt{2} \text{DA}^{\Sigma^0}  \ , \\
  \text{DA}^\Xi &\equiv \text{DA}^{\Xi^0} = - \text{DA}^{\Xi^-}  \ ,
 \end{align}
\end{subequations}
and give the results only for $\text{DA}^N$, $\text{DA}^\Sigma$, $\text{DA}^\Xi$ and $\text{DA}^\Lambda$. In the $\operatorname{SU}(3)_f$ symmetric limit all these DAs can be related to those of the nucleon:%
\begin{subequations}
\begin{align}
 \Phi_{+, i}^\star &\equiv \Phi_{+, i}^{N\star} = \Phi_{+, i}^{\Sigma\star} = \Phi_{+, i}^{\Xi\star} = \Phi_{+, i}^{\Lambda\star} = \Pi_{ i}^{N\star} = \Pi_{i}^{\Sigma\star} = \Pi_{i}^{\Xi\star} \ , \\
 \Phi_{-, i}^\star &\equiv \Phi_{-, i}^{N\star} = \Phi_{-, i}^{\Sigma\star} = \Phi_{-, i}^{\Xi\star} = \Phi_{-, i}^{\Lambda\star} = \Pi_{ i}^{\Lambda\star}\ , \\
 \Xi_{+, i}^\star &\equiv \Xi_{+, i}^{N\star} = \Xi_{+, i}^{\Sigma\star} = \Xi_{+, i}^{\Xi\star} = \Xi_{+, i}^{\Lambda\star}=\Upsilon_{ i}^{N\star} = \Upsilon_{i}^{\Sigma\star} = \Upsilon_{ i}^{\Xi\star} \ , \\
 \Xi_{-, i}^\star &\equiv \Xi_{-, i}^{N\star} = \Xi_{-, i}^{\Sigma\star} = \Xi_{-, i}^{\Xi\star} = \Xi_{-, i}^{\Lambda\star}=\Upsilon_{ i}^{\Lambda\star} \ .
\end{align}
\end{subequations}
\subsection{Minimal parametrization of baryon octet distribution amplitudes} \label{sect_minimal_parametrization}
The choice of DAs presented in the previous section allows us to write down our results in a very compact form:%
\begin{subequations} \label{DAs_chpt_Result}
\begin{align}
 \Phi_{\pm, i}^B &= \Phi_{\pm, i}^\star \Bigl( 1 + \tfrac{1}{2} \Delta\Sigma^\prime_B +  \Delta g^B_{\Phi \pm} \Bigr) + \delta m \  \Delta \Phi_{\pm,i}^B \ , \\*
 \Xi_{\pm, i}^B &= \Xi_{\pm, i}^\star \Bigl( 1 + \tfrac{1}{2} \Delta\Sigma^\prime_B +  \Delta g^B_{\Xi} \Bigr) + \delta m \  \Delta \Xi_{\pm,i}^B \ , \\
 \Pi_{i}^B &=  \Phi_{\pm_B, i}^\star \Bigl( 1 + \tfrac{1}{2} \Delta\Sigma^\prime_B +  \Delta g^B_{\Pi} \Bigr) + \delta m \  \Delta \Pi_{i}^B \ , \\
 \Upsilon_{i}^B &= \Xi_{\pm_B, i}^\star \Bigl( 1 + \tfrac{1}{2} \Delta\Sigma^\prime_B +  \Delta g^B_{\Xi} \Bigr) + \delta m \  \Delta \Upsilon_{i}^B \ ,
\end{align}
\end{subequations}
where ``$\pm_B$'' stands for ``$+$'' if $B\neq\Lambda$ and for ``$-$'' if $B=\Lambda$. The second term in the equations above originates from quark mass insertions, while the first term (or, to be more precise, $\Delta\Sigma^\prime_B$ and $\Delta g^B_{\text{DA}}$) is generated by meson loops and contains chiral logarithms. Owing to our choice of DAs the functions $\Delta g^B_{\text{DA}}$, which are listed in appendix~\ref{sect_gfunctions} together with $\Delta\Sigma^\prime_B$, do not depend on the twist of the amplitude. $\Delta g^B_{\text{DA}}$ and $\Delta\Sigma^\prime_B$ vanish for equal quark masses ($\delta m = 0$). The nontrivial dependence on the mean quark mass of the distribution amplitudes $\Phi_{\pm, i}^\star$ and $\Xi_{\pm, i}^\star$ is presented in section~\ref{sect_mean_quark}. The amplitudes describing the tree-level contribution to the $\operatorname{SU}(3)_f$ symmetry breaking are not completely free. It holds for all distribution amplitudes
\begin{align} \label{DeltaXi}
 \Delta \text{DA}^{\Xi} = - \Delta \text{DA}^{N}- \Delta \text{DA}^{\Sigma} \ .
\end{align}
Furthermore, the amplitudes $\Delta \Pi_i^B$ and $\Delta \Upsilon_i^B$ can be expressed in terms of $\Delta \Phi_{\pm, i}^B$ and $\Delta \Xi_{\pm, i}^B$:%
\begin{subequations}\label{constraints_on_Delta_DAs}
\begin{align}
  \Delta \Pi_i^{N} &= \Delta \Phi_{+,i}^{N} \ , &   
  \Delta \Upsilon_i^{N} &= \Delta \Xi_{+,i}^{N} \ , \\*
  \Delta \Pi_i^{\Sigma} &= -\frac{1}{2} \Delta \Phi_{+,i}^{\Sigma} -\frac{3}{2} \Delta \Phi_{+,i}^{\Lambda} \ , &
  \Delta \Upsilon_i^{\Sigma} &= -\frac{1}{2}\Delta \Xi_{+,i}^{\Sigma} - \frac{3}{2} \Delta \Xi_{+,i}^{\Lambda} \ , \\*
  \Delta \Pi_i^{\Lambda} &= -\frac{1}{2} \Delta \Phi_{-,i}^{\Lambda} -\frac{3}{2} \Delta \Phi_{-,i}^{\Sigma} \ , &
  \Delta \Upsilon_i^{\Lambda} &= -\frac{1}{2}\Delta \Xi_{-,i}^{\Lambda} - \frac{3}{2} \Delta \Xi_{-,i}^{\Sigma} \ ,
\end{align}
\end{subequations}%
which means that the $\Pi^B_i$ and $\Upsilon^B_i$ are completely fixed by the other amplitudes. The divergencies of leading one-loop order contained in $\Delta \Sigma^\prime_B$ and $\Delta g^B_{\text{DA}}$ can be canceled by the introduction of counterterms%
\begin{align}\label{counterterms}
 \Delta \Phi^B_{\pm,i} &\longrightarrow  \frac{{m_b^\star}^2 c^B_{\Phi \pm}}{24 F_\star^2} \Phi^{\star}_{\pm,i} L + \Delta \Phi^{B,\text{ren.}}_{\pm,i}(\mu) \ , &
 \Delta \Xi^B_{\pm,i} &\longrightarrow  \frac{{m_b^\star}^2 c^B_{\Xi}}{24 F_\star^2} \Xi^{\star}_{\pm,i} L + \Delta \Xi^{B,\text{ren.}}_{\pm,i}(\mu) \ ,
\end{align}
where $L$ contains the divergence and the typical constants of the modified minimal subtraction scheme (see eq.~\eqref{def_divergence}). $F_\star$ is the meson decay constant in the $\operatorname{SU(3)}_f$ symmetric limit. The coefficients $c^B_{\text{DA}}$ are given by%
\begin{align}
c^{N}_{\Phi \pm} &= -9 (D^2+10 D F - 3 F^2) -23 \mp 24  \ , &
c^{N}_{\Xi} &= -9 (D^2+10 D F - 3 F^2) +9 \ , \nonumber \\*
c^{\Sigma}_{\Phi \pm} &= 18 (D^2 - 3 F^2) +10 \pm 12 \ , &
 c^{\Sigma}_{\Xi} &= - c^{\Lambda}_{\Xi}= 18 (D^2 - 3 F^2) -18 \ , \nonumber \\*
c^{\Lambda}_{\Phi \pm} &= -18 (D^2 - 3 F^2) -26 \pm 12 \ .
\end{align}
Note that we give no values for $c^{\Xi}_{\Phi \pm}$ and $c^{\Xi}_{\Xi}$, since the renormalization of the corresponding amplitudes is already fixed via eq.~\eqref{DeltaXi}. The renormalized amplitudes acquire a dependence on the chiral renormalization scale $\mu$, which exactly cancels the scale dependence of the leading chiral logarithms:%
\begin{align}
\mu \frac{\partial}{\partial \mu}  \Delta \Phi^{B,\text{ren.}}_{\pm,i}(\mu) &= \frac{-1}{(4\pi)^2} \frac{{m_b^\star}^2 c^B_{\Phi \pm}}{24 F_\star^2} \Phi^\star_{\pm,i} \ , &
\mu \frac{\partial}{\partial \mu}  \Delta \Xi^{B,\text{ren.}}_{\pm,i}(\mu) &= \frac{-1}{(4\pi)^2} \frac{{m_b^\star}^2 c^B_{\Xi}}{24 F_\star^2} \Xi^\star_{\pm,i} \ .
\end{align}
The replacements given in eq.~\eqref{counterterms} also have to cancel the divergencies in the distribution amplitudes for the $\Xi$ baryon and the $\Pi^B_i$ and $\Upsilon^B_i$ distribution amplitudes, which is the case and can be seen as a nontrivial check of our calculation. The higher order divergencies, which are contained in our result as a consequence of using $\text{IR}$-regularization~\cite{Becher:1999he}, have to be set to zero by hand. This introduces an unphysical scale dependence in higher order terms, which is usually solved by fixing the scale at a typical hadronic value like $\unit{1}{\giga\electronvolt}$. A variation of this scale within reasonable bounds, say between $\unit{0.8}{\giga\electronvolt}$ and $\unit{1.2}{\giga\electronvolt}$, can be used to estimate higher order effects. \par
If we neglect higher order contributions, we can rewrite eqs.~\eqref{DAs_chpt_Result} in such a way that the complete nonanalytic behaviour is encoded in an overall prefactor:%
\begin{subequations} \label{DAs_chpt_Result_factorized}
\begin{align}
\begin{split}
 \Phi_{\pm, i}^B  &\mathrel{\dot =} \sqrt{\frac{Z_B}{Z^\star}}\Bigl( 1 +  \Delta g^B_{\Phi \pm} \Bigr) \Bigl( \Phi_{\pm, i}^\star  + \delta m \  \Delta \Phi_{\pm,i}^B \Bigr)  \ , 
\end{split} \\
\begin{split}
 \Xi_{\pm, i}^B &\mathrel{\dot =} \sqrt{\frac{Z_B}{Z^\star}} \Bigl( 1  +  \Delta g^B_{\Xi} \Bigr) \Bigl(\Xi_{\pm, i}^\star  + \delta m \  \Delta \Xi_{\pm,i}^B \Bigr) \ ,
\end{split} \\
\begin{split}
 \Pi_{i}^B &\mathrel{\dot =} \sqrt{\frac{Z_B}{Z^\star}} \Bigl( 1 +  \Delta g^B_{\Pi} \Bigr) \Bigl( \Phi_{\pm_B, i}^\star + \delta m \  \Delta \Pi_{i}^B \Bigr)\ ,
\end{split} \\
\begin{split}
  \Upsilon_{i}^B &\mathrel{\dot =} \sqrt{\frac{Z_B}{Z^\star}} \Bigl( 1 +   \Delta g^B_{\Xi} \Bigr) \Bigl( \Xi_{\pm_B, i}^\star  + \delta m \  \Delta \Upsilon_{i}^B \Bigr) \ ,
\end{split}
\end{align}
\end{subequations}
where%
\begin{align}
 \sqrt{\frac{Z_B}{Z^\star}} &\mathrel{\dot =} 1+\frac{1}{2}\Delta \Sigma^\prime_B \ .
\end{align}
From eq.~\eqref{DAs_chpt_Result_factorized} it follows directly that at leading one-loop order the complete nonanalytic structure is contained in the normalization of the distribution amplitudes, while their shape only exhibits the simple dependence on $\delta m$ shown in eq.~\eqref{shape_of_DAs}. Therefore leading finite volume effects do only affect the normalization. We want to emphasize that this is only true by virtue of our specific choice of DAs. A similar behaviour was found for the meson sector (see refs.~\cite{Chen:2003fp,Chen:2005js}). The zeroth moments of the given DAs are not independent, due to eq.~\eqref{zeroth_moments}. In particular all DAs which correspond to operators of certain symmetry classes are normalized by the same wave function normalization constants independent of the twist of the corresponding amplitude. The zeroth moments define the following normalization constants:%
\begin{subequations} \label{normalization_constants}
\begin{align}
f^B &=\int [dx]\Phi^B_{+,i}(x_1,x_2,x_3) = \sqrt{\frac{Z_B}{Z^\star}}\Bigl( 1 +  \Delta g^B_{\Phi +} \Bigr) \Bigl(f^\star  + \delta m \  \Delta f^B \Bigr) \ , \\
\lambda_1^B &= \int [dx]\Phi^B_{-,4/5}(x_1,x_2,x_3) = \sqrt{\frac{Z_B}{Z^\star}}\Bigl( 1 +  \Delta g^B_{\Phi -} \Bigr) \Bigl(\lambda_1^\star  + \delta m \  \Delta \lambda_1^B \Bigr) \ , \\
\lambda_2^B &= \int [dx]\Xi^B_{+,4/5}(x_1,x_2,x_3) = \sqrt{\frac{Z_B}{Z^\star}}\Bigl( 1 +  \Delta g^B_{\Xi} \Bigr) \Bigl(\lambda_2^\star  + \delta m \  \Delta \lambda_2^B \Bigr) \ ,
\shortintertext{and}
f_T^\Sigma&=\int [dx]\Pi^{\Sigma}_{i}(x_1,x_2,x_3) = \sqrt{\frac{Z_\Sigma}{Z^\star}}\Bigl( 1 +  \Delta g^\Sigma_{\Pi} \Bigr) \Bigl(f^\star  + \delta m \  \Delta f_T^\Sigma \Bigr) \ , \\
f_T^\Xi&=\int [dx]\Pi^{\Xi}_{i}(x_1,x_2,x_3) = \sqrt{\frac{Z_\Xi}{Z^\star}}\Bigl( 1 +  \Delta g^\Xi_{\Pi} \Bigr) \Bigl(f^\star  + \delta m \ \Delta f_T^\Xi \Bigr) \ , \\
\lambda_T^\Lambda&=\int [dx]\Pi^{\Lambda}_{4/5}(x_1,x_2,x_3) = \sqrt{\frac{Z_\Lambda}{Z^\star}}\Bigl( 1 +  \Delta g^\Lambda_{\Pi} \Bigr) \Bigl(\lambda_1^\star  + \delta m \  \Delta \lambda_T^\Lambda \Bigr) \ .
\intertext{For the remaining zeroth moments one finds}
f^{N} &=\int [dx]\Pi^{N}_{i}(x_1,x_2,x_3) \ , \\
0&=\int [dx]\Phi^B_{-,3/6}(x_1,x_2,x_3) = \int [dx]\Xi^B_{-,4/5}(x_1,x_2,x_3) = \int [dx]\Pi^{\Lambda}_{3/6}(x_1,x_2,x_3)   \ , \\
& \mathrel{\phantom{=}} \int [dx]\Upsilon^{B}_{4/5}(x_1,x_2,x_3) = \begin{cases} \lambda_2^B \ , &\text{for } B \neq \Lambda \\ 0 \ , &\text{for } B = \Lambda  \end{cases} \ .
\end{align}
\end{subequations}
Due to eq.~\eqref{constraints_on_Delta_DAs},%
\begin{align} \label{Delta_fT_LambdaT}
 \Delta f_T^\Sigma &= -\frac{3}{2}\Delta f^\Lambda - \frac{1}{2} \Delta f^\Sigma \ , & 
\Delta f_T^\Xi &= \frac{3}{2}\Delta f^\Lambda +\frac{1}{2}\Delta f^\Sigma-\Delta f^{N} \ , \nonumber \\ 
\Delta \lambda_T^\Lambda &= -\frac{1}{2}\Delta \lambda_1^\Lambda -\frac{3}{2} \Delta \lambda_1^\Sigma \ .
\end{align}
In the equations above we have introduced convenient new definitions of $f^\Lambda$, $\lambda_1^\Lambda$, $\lambda_2^\Lambda$, $f_T^\Sigma$, $f_T^\Xi$ and $\lambda_T^\Lambda$ such that, in the limit of exact $SU(3)_f$ symmetry,%
\begin{subequations}
\begin{align}
f^\star &= f^N = f^\Sigma = f^\Xi = f^\Lambda = f^\Sigma_T = f^\Xi_T \ , \\*
\lambda_1^\star &= \lambda_1^N = \lambda_1^\Sigma = \lambda_1^\Xi = \lambda_1^\Lambda = \lambda_T^\Lambda , \\*
\lambda_2^\star &= \lambda_2^N = \lambda_2^\Sigma = \lambda_2^\Xi = \lambda_2^\Lambda \ .
\end{align}
\end{subequations}
If the reader favors a different definition he or she can easily read off the conversion factor from eq.~\eqref{definition_superior_DAs}, noting that additional signs can arise from eq.~\eqref{baryon_signs} if one uses different baryons for the definition of the distribution amplitudes, and that one has to take into account additional factors originating from differing definitions of $S_i$, $P_i$, $V_i$, $A_i$ and $T_i$ (we use the definitions of ref.~\cite{Braun:2000kw}). We have performed this matching procedure for the constants defined in refs.~\cite{Chernyak:1987nu,Braun:2000kw} (see appendix~\ref{app_matching}). Note that the constants $f_T^\Sigma$, $f_T^\Xi$ and $\lambda_T^\Lambda$ given above are (at leading one-loop accuracy) completely fixed by $f^B$ and $\lambda_1^B$. However, without the knowledge of the $\operatorname{SU}(3)_f$ breaking effects one would have to define them as additional free normalization constants. $f^\star$, $\Delta f^B$, $\lambda_i^\star$ and $\Delta \lambda_i^B$ are given by%
\begin{subequations}
\begin{align}
 f^\star &= \int [dx]\Phi^\star_{+,i}(x_1,x_2,x_3) \ , &  \Delta f^B &= \int [dx]\Delta \Phi^B_{+,i}(x_1,x_2,x_3) \ , \\
 \lambda_1^\star &= \int [dx]\Phi^\star_{-,4/5}(x_1,x_2,x_3) \ , &  \Delta \lambda_1^B &= \int [dx]\Delta \Phi^B_{-,4/5}(x_1,x_2,x_3) \ ,  \\
 \lambda_2^\star &= \int [dx]\Xi^\star_{+,4/5}(x_1,x_2,x_3) \ , &  \Delta \lambda_2^B &= \int [dx]\Delta \Xi^B_{+,4/5}(x_1,x_2,x_3) \ ,
\end{align}
\end{subequations}
where, as a consequence of eq.~\eqref{DeltaXi} (first line) and eq.~\eqref{zeroth_moments} (second line) one has%
\begin{align}
 \Delta f^\Xi &=  - \Delta f^\Sigma - \Delta f^N \ , & 
\Delta \lambda_1^\Xi &=  - \Delta \lambda_1^\Sigma - \Delta \lambda_1^N \ ,  &
\Delta \lambda_2^\Xi &=  - \Delta \lambda_2^\Sigma - \Delta \lambda_2^N \ , \nonumber \\*
\Delta \lambda_2^\Lambda &=  - \Delta \lambda_2^\Sigma \ . \label{Delta_f_Lambda}
\end{align}
The zeroth moments of $\Phi^B_{-,3/6}$ and $\Pi^\Lambda_{3/6}$ ($\Xi^B_{-,4/5}$ and $\Upsilon^\Lambda_{4/5}$) vanish by construction, since they are antisymmetric under exchange of $x_1$ and $x_3$ ($x_2$ and $x_3$). One possible approach would be to normalize these amplitudes by their first moments. However, our main goal is to divide the DAs by normalization constants in such a way that the nonanalytic prefactor is canceled. This can be achieved without the definition of additional constants, since all prefactors present in eqs.~\eqref{DAs_chpt_Result_factorized} also occur in eqs.~\eqref{normalization_constants}. Explicitly, one can consider the ratios%
\begin{subequations} \label{shape_of_DAs}
 \begin{align}
\frac{\Phi_{+, i}^B}{f^B}&= \frac{\Phi_{+, i}^\star  + \delta m \  \Delta \Phi_{+,i}^B}{f^\star  + \delta m \  \Delta f^B} \ , &
\frac{\Phi_{-, i}^B}{\lambda_1^B}&= \frac{\Phi_{-, i}^\star  + \delta m \  \Delta \Phi_{-,i}^B}{\lambda_1^\star  + \delta m \  \Delta \lambda_1^B} \ , \\*
\frac{\Pi_i^{N}}{f^{N}}&= \frac{\Phi_{+, i}^\star  + \delta m \  \Delta \Phi_{+,i}^{N}}{f^\star  + \delta m \  \Delta f^{N}} \ ,
& \frac{\Pi_i^\Lambda}{\lambda^\Lambda_T}&= \frac{\Phi_{-, i}^\star  + \delta m \  \Delta \Pi_i^\Lambda}{\lambda_1^\star  + \delta m \  \Delta \lambda_T^\Lambda} \ , \\*
\frac{\Pi_i^{\Sigma/\Xi}}{f^{\Sigma/\Xi}_T}&= \frac{\Phi_{+, i}^\star  + \delta m \  \Delta \Pi_i^{\Sigma/\Xi}}{f^\star  + \delta m \  \Delta f_T^{\Sigma/\Xi}} \ ,  \\
    \frac{\Xi_{\pm, i}^B}{\lambda_2^B}&= \frac{\Xi_{\pm, i}^\star  + \delta m \  \Delta \Xi_{\pm, i}^B}{\lambda_2^\star  + \delta m \  \Delta \lambda_2^B} \ , & \frac{\Upsilon_i^B}{\lambda_2^B}&=  \frac{\Xi_{\pm_B, i}^\star  + \delta m \  \Delta \Upsilon_i^B}{\lambda_2^\star  + \delta m \  \Delta \lambda_2^B}  \ .
 \end{align}
\end{subequations}
The idea behind the latter choice is to normalize all DAs with similar behaviour under chiral extrapolation (including the ones with vanishing zeroth moment) with the same normalization constant containing the complete nonanalytic behaviour. In this way one obtains a one-to-one correspondence between a normalization constant and a certain chiral behaviour. Note that, following this argument, some of the moments of the leading twist DA $\Phi_3^B = \Phi_{+,3}^B + \Phi_{-,3}^B$ should be normalized with $\lambda_1^B$ instead of $f^B$. Otherwise the corresponding shape parameters do contain chiral logarithms.%
\subsection{Example of application} \label{sect_example_of_application}
In this section we will work out explicit expressions for the DAs defined in eqs.~\eqref{definition_superior_DAs} and~\eqref{definition_superior_DAs_second_part} in terms of the shape parameters given in refs.~\cite{Anikin:2013aka,Anikin:2013yoa,Braun:2008ia}, where contributions of Wandzura-Wilczek type~\cite{Wandzura:1977qf} are taken into account explicitly. For brevity we apply the approximation advocated in ref.~\cite{Anikin:2013aka}, where contributions that can mix with four-particle operators are systematically neglected. We use the definitions of said references and we define additionally%
\begin{align} \label{parity_polynomials}
 \mathcal P_{nk}(x_1,x_2,x_3)&= p_{nk} \mathcal P_{nk}(x_3,x_2,x_1) \ ,
\end{align}
where $p_{nk}=\pm1$, depending on $n$ and $k$. This definition is possible since the polynomials $\mathcal P_{nk}$ have definite parity under exchange of $x_1$ and $x_3$ \cite{Anikin:2013aka}. We will call the polynomials with $p_{nk}=+1$ ($p_{nk}=-1$) even (odd). For the DAs we find%
\begin{subequations}  \label{new_def}
 \begin{align}
\Phi^B_{+,3} &= f^B  \Phi_{+,3}^{B,t=3}\ , \\*
\Phi^B_{-,3}&= \lambda_1^B \Phi_{-,3}^{B,t=3} \ , \\*
\Phi^B_{+,4} &= f^B \Bigl( \Phi_{+,4}^{B,WW_3} + \Phi_{+,4}^{B,t=4} \Bigr) \ , &
\Xi^B_{\pm,4} &= \lambda_2^B  \Xi_{\pm,4}^{B,t=4} \ , \\*
\Phi^B_{-,4} &= \lambda_1^B \Bigl(\Phi_{-,4}^{B,WW_3} + \Phi_{-,4}^{B,t=4}  \Bigr) \ , \\
 \Phi^B_{+,5} &= f^B \Bigl( \Phi_{+,5}^{B,WW_3} + \Phi_{+,5}^{B,WW_4} + \Phi_{+,5}^{B,t=5}  \Bigr) \ , &
 \Xi^B_{\pm,5} &= \lambda_2^B  \Bigl( \Xi_{\pm,5}^{B,WW_4} + \Xi_{\pm,5}^{B,t=5}  \Bigr) \ , \\*
 \Phi^B_{-,5} &= \lambda_1^B \Bigl(  \Phi_{-,5}^{B,WW_3} + \Phi_{-,5}^{B,WW_4} + \Phi_{-,5}^{B,t=5}   \Bigr) \ ,
\end{align}
\end{subequations}
where all chiral logarithms are contained in the prefactors. Analogous expressions for the $\Pi$ and $\Upsilon$ DAs will be given below in eq.~\eqref{new_def_2}. Genuine twist $5$ contributions ($\Phi_{\pm,5}^{B,t=5}$, $\Xi_{\pm,5}^{B,t=5}$) will be neglected in this approximation. Also twist $6$ DAs are neglected; one could in principle take into account Wandzura-Wilczek contributions to the twist $6$ DAs, but the corresponding expressions are not known yet. The shape of the DAs is given by the genuine twist $3$ and twist $4$ contributions%
\begin{subequations}  
\begin{align}
 \Phi_{+,3}^{B,t=3}(x_1,x_2,x_3) &= 120 x_1 x_2 x_3 \sum_{\mathclap{\stackrel{n,k\leq n}{p_{nk}=+1}}} \varphi^B_{nk} \mathcal P_{nk} (x_1,x_2,x_3) \ , \\*
  \Phi_{-,3}^{B,t=3}(x_1,x_2,x_3) &= 120 x_1 x_2 x_3 \sum_{\mathclap{\stackrel{n,k\leq n}{p_{nk}=-1}}} \tilde \varphi^B_{nk} \mathcal P_{nk} (x_1,x_2,x_3) \ , \\
\begin{split}
\Phi_{+,4}^{B,t=4}(x_1,x_2,x_3) &=  24  x_1 x_2 \biggl(  \frac{10}{3} (2 x_1 - x_2 - 2 x_3) \tilde \eta^B_{11}  + \dots \biggr)\ ,
\end{split} \\
\begin{split}
\Phi_{-,4}^{B,t=4}(x_1,x_2,x_3) &= 24 x_1 x_2 \bigl(  \eta^B_{00} + 2 (2 - 5 x_2) \eta^B_{10} + \dots \bigr) \ ,
\end{split} \\
\begin{split}
\Xi_{+,4}^{B,t=4}(x_1,x_2,x_3) &= 24 x_2 x_3 \biggl(\xi^B_{00} - \frac{9}{4} (1 - 5 x_1) \xi^B_{10} + \dots \biggr) \ ,
\end{split} \\
\begin{split}
\Xi_{-,4}^{B,t=4}(x_1,x_2,x_3) &= 24 x_2 x_3 \biggl(-\frac{45}{4}(x_2-x_3)\xi^B_{10} + \dots \biggr) \ ,
\end{split}
\end{align}
\end{subequations}
and the Wandzura-Wilczek contributions (see refs.~\cite{Anikin:2013aka,Anikin:2013yoa,Braun:2008ia})%
\begin{subequations} \label{WandzuraWilczek} 
\begin{align}
\Phi_{+,4}^{B,WW_3}(x_1,x_2,x_3)&= -\sum_{\mathclap{\stackrel{n,k\leq n}{p_{nk}=+1}}} \frac{240\varphi^B_{nk}}{(n+2)(n+3)} \biggl( n +2 -\frac{\partial}{\partial x_3} \biggr)  x_1 x_2 x_3 \mathcal P_{nk} (x_1,x_2,x_3)\ , \\*
 \Phi_{-,4}^{B,WW_3}(x_1,x_2,x_3) &= - \sum_{\mathclap{\stackrel{n,k\leq n}{p_{nk}=-1}}} \frac{240 \tilde \varphi^B_{nk}}{(n+2)(n+3)} \biggl( n +2 -\frac{\partial}{\partial x_3} \biggr)  x_1 x_2 x_3  \mathcal  P_{nk} (x_1,x_2,x_3) \ , \\
\begin{split}
\Phi_{+,5}^{B,WW_3}(x_1,x_2,x_3)&= \sum_{\mathclap{\stackrel{n,k\leq n}{p_{nk}=+1}}} \begin{aligned}[t] & \frac{240\varphi^B_{nk}}{(n+2)(n+3)} \biggl[\biggl( n +2 -\frac{\partial}{\partial x_1} \biggr)\biggl( n +1 -\frac{\partial}{\partial x_2}\biggr)  - (n+2)^2 \biggr]\\ &\quad \times x_1 x_2 x_3 \mathcal P_{nk} (x_1,x_2,x_3) \ , \end{aligned}
\end{split} \\
\begin{split}
\Phi_{-,5}^{B,WW_3}(x_1,x_2,x_3)&= \sum_{\mathclap{\stackrel{n,k\leq n}{p_{nk}=-1}}} \begin{aligned}[t] & \frac{240\tilde\varphi^B_{nk}}{(n+2)(n+3)} \biggl[\biggl( n +2 -\frac{\partial}{\partial x_1} \biggr)\biggl( n +1 -\frac{\partial}{\partial x_2}\biggr) - (n+2)^2  \biggr]\\ &\quad \times x_1 x_2 x_3 \mathcal P_{nk} (x_1,x_2,x_3) \ , \end{aligned}
\end{split} \\
\begin{split}
\Phi_{+,5}^{B,WW_4}(x_1,x_2,x_3)  &=  4  x_3 \bigl(  5 (x_1^2 + 2 x_2 x_3 - x_3^2) \tilde \eta^B_{11}  + \dots \bigr)\ ,
\end{split} \\
\begin{split}
\Phi_{-,5}^{B,WW_4}(x_1,x_2,x_3) &= 4 x_3 (1 - x_2) \bigl( 2  \eta^B_{00} + 3 (1 - 5 x_2) \eta^B_{10} + \dots \bigr) \ ,
\end{split} \\
\begin{split}
\Xi_{+,5}^{B,WW_4}(x_1,x_2,x_3) 
&=   4 x_1 (1 + x_1) \xi^B_{00} - \frac{27}{2} (4 - 4 x_1 + x_1^2- 5 x_1^3)  \xi^B_{10} + \dots  \ ,
\end{split} \\
\begin{split}
\Xi_{-,5}^{B,WW_4}(x_1,x_2,x_3) 
&=  - 12 x_1 (x_2 - x_3) \xi^B_{00} + \frac{27}{2}(5-x_1 + 5 x_1^2) (x_2 - x_3)  \xi^B_{10} + \dots  \ ,
\end{split}
\end{align}
\end{subequations}
where the summation over $n$ starts from $0$ and, generally, goes to infinity, but is truncated at $n=2$ in the approximation of ref.~\cite{Anikin:2013aka}.\footnote{We do not take into account possible quark mass corrections to eq.~\eqref{WandzuraWilczek} and eq.~\eqref{WandzuraWilczek2} below (compare e.g.\ refs.~\cite{Ball:1998sk,Braun:2004vf} where such computations have been performed for vector-meson and pseudoscalar-meson DAs), since they can (by definition) be absorbed into the genuine higher twist terms in eq.~\eqref{new_def} and~\eqref{new_def_2}. Let us note in passing that our general result does not rely on the separation of Wandzura-Wilczek and genuine higher twist terms at all, since the calculation within chiral perturbation theory does not distinguish between these contributions.} Note that our separation into ``$+$'' and ``$-$'' amplitudes at leading twist level corresponds to a separation of even and odd polynomials. The normalization constants are still defined such that $\eta^B_{00} = \varphi^B_{00}=\xi^B_{00}=1$, which are only kept for a cleaner notation. Note also that the introduction of $\tilde\varphi^B_{nk}$ and $\tilde\eta^B_{nk}$ only amounts to a redefinition of the shape parameters occurring in $\Phi^{B,t=3}_{-,3}$, $\Phi_{-,4}^{B,WW_3}$ and  $\Phi_{-,5}^{B,WW_3}$ by a factor of $f^B/\lambda_1^B$ and the ones occurring in $  \Phi_{+,4}^{B,t=4}$ and $ \Phi_{+,5}^{B,WW_4}$ by a factor of $\lambda_1^B/f^B$ with respect to ref.~\cite{Anikin:2013aka} (the corresponding anomalous dimensions have to be adjusted accordingly):%
\begin{align}
  \tilde \varphi^B_{nk} &=  \frac{f^B}{\lambda_1^B} \varphi^B_{nk} \ ,  & \tilde \eta^B_{nk} &=  \frac{\lambda_1^B}{f^B} \eta^B_{nk} \ .
\end{align}
The dependence of the shape parameters on the quark mass splitting takes the following form%
\begin{subequations} \label{shape_parameters_1}
\begin{align}
 \varphi^B_{nk} &= \frac{\varphi^\star_{nk}  + \delta m \  \Delta\varphi^B_{nk}}{f^\star  + \delta m \  \Delta f^B} \ , \quad \text{if } p_{nk}=+1 \ , &
 \tilde\varphi^B_{nk} &= \frac{\tilde\varphi^\star_{nk}  + \delta m \  \Delta\tilde\varphi^B_{nk}}{\lambda_1^\star  + \delta m \  \Delta \lambda_1^B} \ , \quad \text{if } p_{nk}=-1 \ , \\
 \tilde\eta^B_{11} &= \frac{ \tilde\eta^\star_{11}  + \delta m \  \Delta \tilde\eta^B_{11}}{f^\star  + \delta m \  \Delta f^B} \ , &  \eta^B_{10} &= \frac{ \eta^\star_{10}  + \delta m \  \Delta \eta^B_{10}}{\lambda_1^\star  + \delta m \  \Delta \lambda_1^B} \ , \\*
 \xi^B_{10} &= \frac{ \xi^\star_{10}  + \delta m \  \Delta \xi^B_{10}}{\lambda_2^\star  + \delta m \  \Delta \lambda_2^B} \ ,
\end{align}
\end{subequations}
which corresponds directly to eq.~\eqref{shape_of_DAs}, while the dependence of the normalization constants is given in eq.~\eqref{normalization_constants}. The parameters describing $\operatorname{SU}(3)_f$ symmetry breaking are restricted by eq.~\eqref{DeltaXi} such that%
\begin{align} \label{su3cons_params_1}
  \Delta x^\Xi_{nk} &= - \Delta x^{N}_{nk} - \Delta x^\Sigma_{nk} \ , \quad \text{for } x \in \{ \varphi,\tilde\varphi,\eta,\tilde\eta,\xi \} \ .
\end{align}
For the original twist $3$ and $4$ DAs given in ref.~\cite{Braun:2000kw} (see also eq.~\eqref{original_DAs_definition}) the new choice of normalization yields%
\begin{subequations}
\begin{align}
  \begin{split}
  \Phi^N_3(x_1,x_2,x_3) &= \Bigl(\Phi^N_{+,3} + \Phi^N_{-,3}\Bigr)(x_1,x_2,x_3) \\*
  &= f^N \Phi^{N,t=3}_{+,3}(x_1,x_2,x_3) + \lambda_1^N \Phi^{N,t=3}_{-,3} (x_1,x_2,x_3) \ ,
\end{split} \\
\begin{split}
 \Phi^N_4(x_1,x_2,x_3) &= \frac{1}{2} \Bigl( \Phi^N_{+,4} + \Phi^N_{-,4} \Bigr)(x_1,x_2,x_3) \\
  &= \frac{f^N}{2} \Bigl( \Phi_{+,4}^{N,WW_3} \! + \Phi_{+,4}^{N,t=4} \Bigr)(x_1,x_2,x_3) +  \frac{\lambda_1^N }{2} \Bigl( \Phi_{-,4}^{N,t=4} + \Phi_{-,4}^{N,WW_3} \! \Bigr)(x_1,x_2,x_3) \ , \!
\end{split} \\
\begin{split}
 \Psi^N_4(x_1,x_2,x_3) &= \frac{1}{2} \Bigl( \Phi^N_{+,4} - \Phi^N_{-,4} \Bigr)(x_3,x_1,x_2)\\
  &= \frac{f^N}{2} \Bigl( \Phi_{+,4}^{N,WW_3} \! + \Phi_{+,4}^{N,t=4} \Bigr)(x_3,x_1,x_2) -  \frac{\lambda_1^N }{2} \Bigl( \Phi_{-,4}^{N,t=4} + \Phi_{-,4}^{N,WW_3} \! \Bigr)(x_3,x_1,x_2)   \ , \!
\end{split} \\
\begin{split}
 \Xi^N_4(x_1,x_2,x_3) &= \frac{1}{6} \Bigl( \Xi^N_{+,4} + \Xi^N_{-,4} \Bigr)(x_1,x_2,x_3)\\
  &= \frac{\lambda_2^N}{6} \Bigl(\Xi_{+,4}^{N,t=4} + \Xi_{-,4}^{N,t=4} \Bigr)(x_1,x_2,x_3)   \ ,
\end{split}
 \end{align}
\end{subequations}
where, as discussed above, the normalization of the odd moments of the leading twist amplitude with $\lambda_1^N$ (instead of $f^N$) appropriately reflects their chiral behaviour. Note that this is consistent with an earlier two-flavor BChPT calculation, where it was found that the odd first and second moments of the leading twist amplitude have the same chiral logarithms as $\lambda_{1}^N$ (see appendix of ref.~\cite{PhysRevD.89.094511}).\par
For a description of the complete baryon octet one also needs the $\Pi$ and $\Upsilon$ DAs defined in eq.~\eqref{definition_superior_DAs_second_part}, which are relevant for the hyperons. These are completely fixed by the $\Phi_\pm$ and $\Xi$ DAs. Consequently, the following equations do not contain any additional parameters:%
\begin{subequations}  \label{new_def_2}
 \begin{align}
\Pi^{N}_{i} &=  \Phi_{+,i}^{N}\ , &
\Upsilon^{N}_{i} &=  \Xi_{+,i}^{N}\ , \\*
\Pi^{\Sigma/\Xi}_{3} &= f^{\Sigma/\Xi}_T  \Pi_{3}^{\Sigma/\Xi,t=3}\ , \\
 \Pi^\Lambda_{3}&= \lambda_T^\Lambda \Pi_{3}^{\Lambda,t=3} \ , \\
 \Pi_{4}^{\Sigma/\Xi} &= f^{\Sigma/\Xi}_T \Bigl( \Pi_{4}^{\Sigma/\Xi,WW_3} + \Pi_{4}^{\Sigma/\Xi,t=4} \Bigr) \ , &
 \Upsilon^B_{4} &= \lambda_2^B  \Upsilon_{4}^{B,t=4} \ , \\
 \Pi_{4}^{\Lambda} &= \lambda_T^\Lambda \Bigl( \Pi_{4}^{\Lambda,WW_3} +  \Pi_{4}^{\Lambda,t=4} \Bigr)\ , \\
 \Pi^{\Sigma/\Xi}_{5} &= f^{\Sigma/\Xi}_T \Bigl( \Pi_{5}^{\Sigma/\Xi,WW_3} + \Pi_{5}^{\Sigma/\Xi,WW_4} + \Pi_{5}^{\Sigma/\Xi,t=5} \Bigr) \ , &
  \Upsilon^B_{5} &= \lambda_2^B  \Bigl( \Upsilon_{5}^{B,WW_4} + \Upsilon_{5}^{B,t=5} \Bigr) \ , \\*
 \Pi^\Lambda_{5} &= \lambda_T^\Lambda \Bigl(  \Pi_{5}^{\Lambda,WW_3} + \Pi_{5}^{\Lambda,WW_4} + \Pi_{5}^{\Lambda,t=5}  \Bigr) \ ,
\end{align}
\end{subequations}
where the genuine twist $5$ contributions $\Pi_{5}^{B,t=5}$ and $\Upsilon_{5}^{B,t=5}$ will be neglected as above. The genuine twist $3$ and twist $4$ contributions are%
\begin{subequations}  
\begin{align}
 \Pi_{3}^{\Sigma/\Xi,t=3}(x_1,x_2,x_3) &= 120 x_1 x_2 x_3 \sum_{\mathclap{\stackrel{n,k\leq n}{p_{nk}=+1}}} \pi^{\Sigma/\Xi}_{nk} \mathcal P_{nk} (x_1,x_2,x_3) \ , \\*
  \Pi_{3}^{\Lambda,t=3}(x_1,x_2,x_3) &= 120 x_1 x_2 x_3 \sum_{\mathclap{\stackrel{n,k\leq n}{p_{nk}=-1}}} \tilde \pi^{\Lambda}_{nk} \mathcal P_{nk} (x_1,x_2,x_3) \ , \\
\begin{split}
\Pi_{4}^{\Sigma/\Xi,t=4}(x_1,x_2,x_3)
 &=  24  x_1 x_2 \biggl(  \frac{10}{3} (2 x_1 - x_2 - 2 x_3) \tilde \zeta^{\Sigma/\Xi}_{11}  + \dots \biggr)\ ,
\end{split} \\
\begin{split}
\Pi_{4}^{\Lambda,t=4}(x_1,x_2,x_3)
 &= 24 x_1 x_2 \bigl(  \zeta^\Lambda_{00} + 2 (2 - 5 x_2) \zeta^\Lambda_{10} + \dots \bigr) \ ,
\end{split} \\
\begin{split}
\Upsilon_{4}^{\Sigma/\Xi,t=4}(x_1,x_2,x_3)
&= 24 x_2 x_3 \biggl(\upsilon^{\Sigma/\Xi}_{00} - \frac{9}{4} (1 - 5 x_1) \upsilon^{\Sigma/\Xi}_{10} + \dots \biggr) \ ,
\end{split} \\*
\begin{split}
\Upsilon_{4}^{\Lambda,t=4}(x_1,x_2,x_3)
&= 24 x_2 x_3 \biggl(-\frac{45}{4}(x_2-x_3)\upsilon^\Lambda_{10} + \dots \biggr) \ .
\end{split}
\end{align}
\end{subequations}
The shape parameters are fixed:%
\begin{subequations} \label{shape_parameters_2}
\begin{align}
 \pi^{\Sigma/\Xi}_{nk} &= \frac{\varphi^\star_{nk}  + \delta m \  \Delta\pi^{\Sigma/\Xi}_{nk}}{f^\star  + \delta m \  \Delta f^{\Sigma/\Xi}_T} \ ,  &
 \tilde\pi^\Lambda_{nk} &= \frac{\tilde\varphi^\star_{nk}  + \delta m \  \Delta\tilde\pi^\Lambda_{nk}}{\lambda_1^\star  + \delta m \  \Delta \lambda_T^\Lambda} \ , \\
 \tilde \zeta^{\Sigma/\Xi}_{11} &= \frac{ \tilde \eta^\star_{11}  + \delta m \  \Delta  \tilde \zeta^{\Sigma/\Xi}_{11}}{f^\star  + \delta m \  \Delta f^{\Sigma/\Xi}_T} \ , &  \zeta^{\Lambda}_{10} &= \frac{ \eta^\star_{10}  + \delta m \  \Delta \zeta^{\Lambda}_{10}}{\lambda_1^\star  + \delta m \  \Delta \lambda_T^\Lambda} \ , \\
 \upsilon^B_{10} &= \frac{ \xi^\star_{10}  + \delta m \  \Delta \upsilon^B_{10}}{\lambda_2^\star  + \delta m \  \Delta \lambda_2^B} \ ,
\end{align}
\end{subequations}
where $ \Delta f^{\Sigma/\Xi}_T$ and $\Delta \lambda_T^\Lambda$ are defined in eq.~\eqref{Delta_fT_LambdaT}. The parameters describing $\operatorname{SU}(3)_f$ symmetry breaking can be determined by eqs.~\eqref{DeltaXi} and~\eqref{constraints_on_Delta_DAs}:%
\begin{subequations} \label{su3cons_params_2}
\begin{align}
\Delta \pi_{nk}^\Sigma &= -\frac{1}{2} \Delta \varphi_{nk}^\Sigma -\frac{3}{2} \Delta \varphi_{nk}^\Lambda \ , & 
\Delta \tilde\pi_{nk}^\Lambda &= -\frac{1}{2} \Delta \tilde\varphi_{nk}^\Lambda -\frac{3}{2} \Delta \tilde\varphi_{nk}^\Sigma \ , \nonumber \\
\Delta \pi_{nk}^\Xi &= \frac{3}{2} \Delta \varphi_{nk}^\Lambda + \frac{1}{2} \Delta \varphi_{nk}^\Sigma  - \Delta \varphi_{nk}^{N} \ , \\
\Delta \tilde\zeta_{11}^\Sigma &= -\frac{1}{2} \Delta \tilde \eta_{11}^\Sigma -\frac{3}{2} \Delta \tilde\eta_{11}^\Lambda \ , &
\Delta \zeta_{10}^\Lambda &= -\frac{1}{2} \Delta \eta_{10}^\Lambda -\frac{3}{2} \Delta \eta_{10}^\Sigma \ , \nonumber \\
\Delta \tilde\zeta_{11}^\Xi &= \frac{3}{2} \Delta \tilde\eta_{11}^\Lambda + \frac{1}{2} \Delta \tilde\eta_{11}^\Sigma  - \Delta \tilde\eta_{11}^{N} \ , \\
 \Delta \upsilon_{10}^\Sigma &= -\frac{1}{2} \Delta \xi_{10}^\Sigma -\frac{3}{2} \Delta \xi_{10}^\Lambda \ ,&
 \Delta \upsilon_{10}^\Lambda &= -\frac{1}{2} \Delta \xi_{10}^\Lambda -\frac{3}{2} \Delta \xi_{10}^\Sigma \ , \nonumber \\
 \Delta \upsilon_{10}^\Xi &= \frac{3}{2} \Delta \xi_{10}^\Lambda + \frac{1}{2} \Delta \xi_{10}^\Sigma  - \Delta \xi_{10}^{N} \ .
\end{align}
\end{subequations}
The Wandzura-Wilczek contributions take the form%
\begin{subequations}\label{WandzuraWilczek2}
\begin{align}
\Pi_{4}^{\Sigma/\Xi,WW_3}(x_1,x_2,x_3)&= -\sum_{\mathclap{\stackrel{n,k\leq n}{p_{nk}=+1}}} \frac{240\pi^{\Sigma/\Xi}_{nk}}{(n+2)(n+3)} \biggl( n +2 -\frac{\partial}{\partial x_3} \biggr)  x_1 x_2 x_3 \mathcal P_{nk} (x_1,x_2,x_3)\ , \\*
 \Pi_{4}^{\Lambda,WW_3}(x_1,x_2,x_3) &= - \sum_{\mathclap{\stackrel{n,k\leq n}{p_{nk}=-1}}} \frac{240 \tilde \pi^\Lambda_{nk}}{(n+2)(n+3)} \biggl( n +2 -\frac{\partial}{\partial x_3} \biggr)  x_1 x_2 x_3  \mathcal  P_{nk} (x_1,x_2,x_3) \ , \\
\begin{split}
\Pi_{5}^{\Sigma/\Xi,WW_3}(x_1,x_2,x_3)&= \sum_{\mathclap{\stackrel{n,k\leq n}{p_{nk}=+1}}} \begin{aligned}[t] & \frac{240\pi^{\Sigma/\Xi}_{nk}}{(n+2)(n+3)} \biggl[\biggl( n +2 -\frac{\partial}{\partial x_1} \biggr)\biggl( n +1 -\frac{\partial}{\partial x_2}\biggr) - (n+2)^2  \biggr]\\ &\quad \times x_1 x_2 x_3 \mathcal P_{nk} (x_1,x_2,x_3) \ , \end{aligned}
\end{split} \\
\begin{split}
\Pi_{5}^{\Lambda,WW_3}(x_1,x_2,x_3)&= \sum_{\mathclap{\stackrel{n,k\leq n}{p_{nk}=-1}}} \begin{aligned}[t] & \frac{240\tilde\pi^\Lambda_{nk}}{(n+2)(n+3)} \biggl[\biggl( n +2 -\frac{\partial}{\partial x_1} \biggr)\biggl( n +1 -\frac{\partial}{\partial x_2}\biggr) - (n+2)^2  \biggr]\\ &\quad \times x_1 x_2 x_3 \mathcal P_{nk} (x_1,x_2,x_3) \ , \end{aligned}
\end{split} \\
\begin{split}
\Pi_{5}^{\Sigma/\Xi,WW_4}(x_1,x_2,x_3) 
 &=  4  x_3 \bigl(  5 (x_1^2 + 2 x_2 x_3 - x_3^2) \tilde \zeta^{\Sigma/\Xi}_{11}  + \dots \bigr)\ ,
\end{split} \\
\begin{split}
\Pi_{5}^{\Lambda,WW_4}(x_1,x_2,x_3)
&= 4 x_3 (1 - x_2) \bigl( 2  \zeta^\Lambda_{00} + 3 (1 - 5 x_2) \zeta^\Lambda_{10} + \dots \bigr) \ ,
\end{split} \\
\begin{split}
\Upsilon_{5}^{\Sigma/\Xi,WW_4}(x_1,x_2,x_3) 
&=  4 x_1 (1 + x_1) \upsilon^{\Sigma/\Xi}_{00} - \frac{27}{2}  (4 - 4 x_1 + x_1^2- 5 x_1^3)  \upsilon^{\Sigma/\Xi}_{10} + \dots  \ ,
\end{split} \\
\begin{split}
\Upsilon_{5}^{\Lambda,WW_4}(x_1,x_2,x_3) 
&=   - 12 x_1 (x_2 - x_3) \upsilon^\Lambda_{00} + \frac{27}{2} (5-x_1 + 5 x_1^2)  (x_2 - x_3)  \upsilon^\Lambda_{10} + \dots \ .
\end{split}
\end{align}
\end{subequations}
To conclude this section we want to point out the merits of our calculation. First of all, we found that the behaviour under chiral extrapolation of a certain moment correlates to its parity in the sense of eq.~\eqref{parity_polynomials}. Therefore it is advantageous to normalize the odd moments of the leading twist DA with $\lambda_1^B$ instead of $f^B$. Quantitatively more important, however, is the significant reduction of parameters: we find that (within the approximation used above) we only need $43$ parameters to describe the complete set of baryon octet three-quark DAs (including their dependence on the quark mass splitting). In contrast, an ad~hoc linear extrapolation without the knowledge of $\operatorname{SU}(3)_f$ symmetry breaking would require $72$ parameters for the given setup, since one can not make use of eqs.~\eqref{Delta_fT_LambdaT}, \eqref{Delta_f_Lambda}, \eqref{su3cons_params_1} and~\eqref{su3cons_params_2}.%
\subsection{Dependence on the mean quark mass} \label{sect_mean_quark}
The distribution amplitudes $\Phi^\star_{\pm,i}$ and $\Xi^\star_{\pm,i}$ have a nontrivial dependence on the mean quark mass $\bar m_q$. This is not really interesting from a phenomenological point of view, since the number of independent distribution amplitudes can not be further reduced compared to eq.~\eqref{DAs_chpt_Result_factorized}, even if one expands everything around the chiral limit. However, the dependence is of importance for the analysis of lattice data if one wants to include data points from simulations with unphysical mean quark mass. The mass dependence reads%
\begin{subequations} \label{DAs_chpt_Result_symmetric_line}
 \begin{align}
 \Phi^\star_{\pm,i} &=  \Phi^\circ_{\pm,i} \Bigl(1+\tfrac{1}{2} \Sigma^{\prime\star} + g^\star_{\Phi\pm} \Bigr) + \bar m \Delta \Phi^\star_{\pm,i} \ , \\
\Xi^\star_{\pm,i} &=  \Xi^\circ_{\pm,i} \Bigl(1+\tfrac{1}{2} \Sigma^{\prime\star} + g^\star_{\Xi} \Bigr) + \bar m \Delta \Xi^\star_{\pm,i} \ ,
\end{align}
\end{subequations}
where%
 \begin{align}
\bar m = \frac{12 B_0 \bar m_q}{{m_b^\star}^2} \ .
\end{align}
$ g^\star_{\Phi\pm}$, $g^\star_{\Xi}$ and $\Sigma^{\prime\star}$ are functions of the mean quark mass that can be taken from appendix~\ref{sect_gfunctions}. The divergencies occurring at linear order in the mean quark mass can be canceled via the following introduction of counterterms%
\begin{align}\label{counterterms_symmetric}
 \Delta \Phi^\star_{\pm,i} &\longrightarrow  \frac{{m_b^\star}^2 c^\star_{\Phi \pm}}{24 F_\star^2} \Phi^{\circ}_{\pm,i} L + \Delta \Phi^{\star,\text{ren.}}_{\pm,i}(\mu) \ , &
 \Delta \Xi^\star_{\pm,i} &\longrightarrow  \frac{{m_b^\star}^2 c^\star_{\Xi}}{24 F_\star^2} \Xi^{\circ}_{\pm,i} L + \Delta \Xi^{\star,\text{ren.}}_{\pm,i}(\mu) \ ,
\end{align}
where $L$ contains the divergence (see appendix~\ref{sect_gfunctions}) and the coefficients are%
\begin{align}
c^{\star}_{\Phi \pm} &= \frac{4}{3} \bigl( 6 (5 D^2 + 9 F^2) + 13 \pm 6 \bigr) \ , &
c^{\star}_{\Xi} &= \frac{4}{3} \bigl( 6 (5 D^2 + 9 F^2) + 9  \bigr)  \ . 
\end{align}
This leads to the following scale dependence in the renormalized amplitudes:%
\begin{align}
\mu \frac{\partial}{\partial \mu}  \Delta \Phi^{\star,\text{ren.}}_{\pm,i}(\mu) &= \frac{-1}{(4\pi)^2} \frac{{m_b^\star}^2 c^\star_{\Phi \pm}}{24 F_\star^2} \Phi^\circ_{\pm,i} \ , &
\mu \frac{\partial}{\partial \mu}  \Delta \Xi^{\star,\text{ren.}}_{\pm,i}(\mu) &= \frac{-1}{(4\pi)^2} \frac{{m_b^\star}^2 c^\star_{\Xi}}{24 F_\star^2} \Xi^\circ_{\pm,i} \ .
\end{align}
The divergencies occurring together with higher orders of the quark masses have to be canceled by hand as discussed in section~\ref{sect_minimal_parametrization}. If one takes eq.~\eqref{DAs_chpt_Result_symmetric_line} and plugs it into eq.~\eqref{DAs_chpt_Result_factorized} one finds (up to terms of higher order)%
\begin{subequations} \label{DAs_chpt_Result_factorized_mean_quark_mass}
\begin{align}
\begin{split}
 \Phi_{\pm, i}^B  &\mathrel{\dot =} \sqrt{Z_B}\Bigl( 1 + g^\star_{\Phi \pm} + \Delta g^B_{\Phi \pm} \Bigr) \Bigl( \Phi^\circ_{\pm,i} +\bar m  \  \Delta\Phi_{\pm, i}^\star  + \delta m \  \Delta \Phi_{\pm,i}^B \Bigr)  \ , 
\end{split} \\*
\begin{split}
 \Xi_{\pm, i}^B  &\mathrel{\dot =} \sqrt{Z_B}\Bigl( 1 + g^\star_{\Xi } + \Delta g^B_{\Xi} \Bigr) \Bigl( \Xi^\circ_{\pm,i} +\bar m \ \Delta\Xi_{\pm, i}^\star  + \delta m \  \Delta \Xi_{\pm,i}^B \Bigr)  \ , 
\end{split} \\
\begin{split}
 \Pi_{i}^B &\mathrel{\dot =}\sqrt{Z_B}   \Bigl( 1 +  g^\star_{\Phi \pm_B} + \Delta g^B_{\Pi} \Bigr) \Bigl( \Phi^\circ_{\pm_B,i} +\bar m  \  \Delta\Phi_{\pm_B, i}^\star + \delta m \  \Delta \Pi_{i}^B \Bigr) \ ,
\end{split} \\*
\begin{split}
  \Upsilon_{i}^B &\mathrel{\dot =} \sqrt{Z_B} \Bigl( 1 + g^\star_{\Xi} +  \Delta g^B_{\Xi} \Bigr) \Bigl( \Xi^\circ_{\pm_B,i} +\bar m  \ \Delta\Xi_{\pm_B, i}^\star  + \delta m \  \Delta \Upsilon_{i}^B \Bigr) \ ,
\end{split}
\end{align}
\end{subequations}
where%
\begin{align}
 \sqrt{Z_B} &\mathrel{\dot =} 1+ \frac{1}{2} \Sigma^{\prime\star}+\frac{1}{2}\Delta \Sigma^\prime_B \ .
\end{align}
Starting from this point everything can be worked out analogously to the case of fixed mean quark mass.%
\section{Summary} \label{sect_summary}
In this work we have presented the first analysis of baryon octet light-cone DAs in the framework of three-flavor BChPT. At next-to-leading order accuracy in the chiral counting scheme, we obtain the leading quark mass dependence and (automatically) the leading $\operatorname{SU}(3)_f$ breaking effects. Describing the baryon octet simultaneously we are able to unify and systemize the efforts made in refs.~\cite{Chernyak:1987nu,Braun:2000kw,Liu:2008yg,Liu:2009uc,Liu:2014uha}.\par
An important insight to be gained from our results is of qualitative nature: in the chiral odd sector the chiral behaviour (i.e.\ the contained chiral logarithms) of a specific moment does not depend on its twist, but on whether it contributes to the $\Phi^B_{+,i}$ or $\Phi^B_{-,i}$ amplitudes (see eq.~\eqref{definition_superior_DAs}). Those contributing to the ``$+$'' (``$-$'') amplitudes have the same chiral logarithms as $f^B$ ($\lambda_1^B$). Therefore the odd moments of the leading twist DA behave like $\lambda_1^B$ instead of (as one might have expected) $f^B$. This result is consistent with an earlier two-flavor calculation, where it was found that the odd first and second moments of the leading twist DA have the same chiral logarithms as $\lambda_{1}^N$ (see appendix of ref.~\cite{PhysRevD.89.094511}).\par
In section~\ref{sect_result} we provide a set of DAs that parametrize the complete baryon octet (including the $\Lambda$ baryon) in a minimal way and do not mix under chiral extrapolation. Eqs.~\eqref{DAs_chpt_Result_factorized} and~\eqref{DAs_chpt_Result_factorized_mean_quark_mass} are our main results. They describe the quark mass dependence of the baryon octet DAs (including all higher twist amplitudes) in a very compact manner. Eq.~\eqref{normalization_constants} contains explicit extrapolation formulas for the wave function normalization constants, while the dependence on the quark mass splitting of the shape parameters, which describe contributions of higher conformal spin, is shown in Eqs.~\eqref{shape_parameters_1} and~\eqref{shape_parameters_2}. The results will be of particular importance for the interpretation and extrapolation of forthcoming lattice QCD data, due to the significant decrease in number of parameters (compare section~\ref{sect_example_of_application}). For the same reason our results are relevant for QCD sum rule analyses and for model building.%
\acknowledgments{We thank P.~C.~Bruns, M.~Gruber, V.~M.~Braun and A.~N.~Manashov for valuable discussions.}
\appendix
\section{Loop contributions} \label{sect_gfunctions}
The functions $g^\star_{\text{DA}}$ and $\Delta g^B_{\text{DA}}$ are given by%
\begin{align}
6 F_\star^2 g_{\Phi +}^{\star}&= -19 \operatorname{H_1}(m_m^\star) + 2 (5 D + 6 F)  \operatorname{H_2}(m_m^\star) \ , \nonumber \\
6 F_\star^2 g_{\Phi -}^{\star}&= -7 \operatorname{H_1}(m_m^\star) - 10  D \operatorname{H_2}(m_m^\star) \ , \nonumber \\
6 F_\star^2 g_{\Xi}^{\star}&= -9 \operatorname{H_1}(m_m^\star) - 18  F \operatorname{H_2}(m_m^\star) \ , \nonumber \\[0.3cm]
24 F_\star^2 \Delta g_{\Phi +}^{N}&=-57 \DeltaHone(m_\pi)-18 \DeltaHone(m_K)-\DeltaHone(m_\eta)+30 (D+F) \DeltaHtwo(m_\pi) \nonumber \\* &\quad +12 (D+F) \DeltaHtwo(m_K)+(-2 D+6 F) \DeltaHtwo(m_\eta)\ , \nonumber \\
24 F_\star^2 \Delta  g_{\Phi +}^{\Sigma}&=-12 \DeltaHone(m_\pi)-60 \DeltaHone(m_K)-4 \DeltaHone(m_\eta)+24 D \DeltaHtwo(m_\pi)\nonumber \\* &\quad +24 (D+2 F) \DeltaHtwo(m_K)-8 D \DeltaHtwo(m_\eta)\ , \nonumber \\
24 F_\star^2 \Delta  g_{\Phi +}^{\Xi}&=-9 \DeltaHone(m_\pi)-66 \DeltaHone(m_K)-\DeltaHone(m_\eta)+18 (-D+F) \DeltaHtwo(m_\pi)\nonumber \\* &\quad +(60 D+36 F) \DeltaHtwo(m_K)-2 (D+3 F) \DeltaHtwo(m_\eta)\ , \nonumber \\
24 F_\star^2 \Delta  g_{\Phi +}^{\Lambda}&=-36 \DeltaHone(m_\pi)-36 \DeltaHone(m_K)-4 \DeltaHone(m_\eta)+24 D \DeltaHtwo(m_\pi)\nonumber \\* &\quad +8 (D+6 F) \DeltaHtwo(m_K)+8 D \DeltaHtwo(m_\eta)\ , \nonumber \\
24 F_\star^2 \Delta  g_{\Phi -}^{N}&=-9 \DeltaHone(m_\pi)-18 \DeltaHone(m_K)-\DeltaHone(m_\eta)-18 (D+F) \DeltaHtwo(m_\pi)\nonumber \\* &\quad +(-20 D+12 F) \DeltaHtwo(m_K)+(-2 D+6 F) \DeltaHtwo(m_\eta)\ , \nonumber \\
24 F_\star^2 \Delta  g_{\Phi -}^{\Sigma}&=-12 \DeltaHone(m_\pi)-12 \DeltaHone(m_K)-4 \DeltaHone(m_\eta)-8 D \DeltaHtwo(m_\pi)\nonumber \\* &\quad -24 D \DeltaHtwo(m_K)-8 D \DeltaHtwo(m_\eta)\ , \nonumber \\
24 F_\star^2 \Delta  g_{\Phi -}^{\Xi}&=-9 \DeltaHone(m_\pi)-18 \DeltaHone(m_K)-\DeltaHone(m_\eta)+18 (-D+F) \DeltaHtwo(m_\pi)\nonumber \\* &\quad -4 (5 D+3 F) \DeltaHtwo(m_K)-2 (D+3 F) \DeltaHtwo(m_\eta)\ , \nonumber \\
24 F_\star^2 \Delta  g_{\Phi -}^{\Lambda}&=-36 \DeltaHone(m_\pi)+12 \DeltaHone(m_K)-4 \DeltaHone(m_\eta)-72 D \DeltaHtwo(m_\pi)\nonumber \\* &\quad +24 D \DeltaHtwo(m_K)+8 D \DeltaHtwo(m_\eta)\ , \nonumber \\
\Delta  g_{\Pi}^{N}&=\Delta  g_{\Phi +}^{N} \ , \nonumber \\
24 F_\star^2 \Delta  g_{\Pi}^{\Sigma}&=-24 \DeltaHone(m_\pi)-36 \DeltaHone(m_K)-16 \DeltaHone(m_\eta)+48 F \DeltaHtwo(m_\pi)\nonumber \\* &\quad +24 D \DeltaHtwo(m_K)+16 D \DeltaHtwo(m_\eta)\ , \nonumber \\
24 F_\star^2 \Delta  g_{\Pi}^{\Xi}&=-9 \DeltaHone(m_\pi)-42 \DeltaHone(m_K)-25 \DeltaHone(m_\eta)+18 (D-F) \DeltaHtwo(m_\pi)\nonumber \\* &\quad +12 (D+3 F) \DeltaHtwo(m_K)+10 (D+3 F) \DeltaHtwo(m_\eta)\ , \nonumber \\
24 F_\star^2 \Delta  g_{\Pi}^{\Lambda}&=-12 \DeltaHone(m_K)-16 \DeltaHone(m_\eta)-24 D \DeltaHtwo(m_K)-16 D \DeltaHtwo(m_\eta)\ , \nonumber \\
24 F_\star^2 \Delta  g_{\Xi}^{N}&=-9 \DeltaHone(m_\pi)-18 \DeltaHone(m_K)-9 \DeltaHone(m_\eta)-18 (D+F) \DeltaHtwo(m_\pi)\nonumber \\* &\quad +12 (D-3 F) \DeltaHtwo(m_K)+6 (D-3 F) \DeltaHtwo(m_\eta)\ , \nonumber \\
24 F_\star^2 \Delta  g_{\Xi}^{\Sigma}&=-24 \DeltaHone(m_\pi)-12 \DeltaHone(m_K)-48 F \DeltaHtwo(m_\pi)-24 F \DeltaHtwo(m_K)\ , \nonumber \\
24 F_\star^2 \Delta  g_{\Xi}^{\Xi}&=-9 \DeltaHone(m_\pi)-18 \DeltaHone(m_K)-9 \DeltaHone(m_\eta)+18 (D-F) \DeltaHtwo(m_\pi)\nonumber \\* &\quad -12 (D+3 F) \DeltaHtwo(m_K)-6 (D+3 F) \DeltaHtwo(m_\eta)\ , \nonumber \\
24 F_\star^2 \Delta  g_{\Xi}^{\Lambda}&=-36 \DeltaHone(m_K)-72 F \DeltaHtwo(m_K) \ .
\end{align}
The $Z$-factor contributions are given by%
\begin{align}
\Sigma^{\prime\star} &= \frac{4}{3}(5D^2+9F^2) \operatorname{H_3}(m_m^\star) \ , \nonumber \\
\Delta \Sigma^\prime_B &= 3 g_{B,\pi} \DeltaHthree(m_\pi) + 4 g_{B,K} \DeltaHthree(m_K) + g_{B,\eta} \DeltaHthree(m_\eta) \ ,
\end{align}
where the coefficients $g_{B,M}$ are defined in eq.~\eqref{def_g_cefficients}. The auxiliary functions $\operatorname{\Delta H_k}$ are defined as%
\begin{align}
\operatorname{\Delta H_k}(m) &= \operatorname{H_k}(m) - \operatorname{H_k}(m_m^\star)  \ ,
\end{align}
with%
\begin{subequations}
\begin{align} 
\operatorname{H_1}(m) &= 2 m^2 \left[ L + \frac{1}{32 \pi^2} \log\left( \frac{m^2}{\mu^2} \right) \right] \ , \\
\operatorname{H_2}(m) &= \frac{m^4}{{m_b^\star}^2} \left[ L + \frac{1}{32 \pi^2} \log\left( \frac{m^2}{\mu^2} \right) \right] - \frac{m^4}{32 \pi^2 {m_b^\star}^2} + \frac{m^3 }{8 \pi^2 m_b^\star} \sqrt{1-\frac{m^2}{4 {m_b^\star}^2}}  \arccos \left(  - \frac{m}{2 m_b^\star } \right) \ , \\
\begin{split}
\operatorname{H_3}(m) &= - \frac{3 m^2}{2 F_\star^2} \left( 1 - \frac{2 m^2}{3 {m_b^\star}^2} \right) \left[ L + \frac{1}{32 \pi^2} \log\left( \frac{m^2}{\mu^2} \right) \right] - \frac{m^2}{32 F_\star^2 \pi^2} \\
&\quad  + \frac{3 m^3 }{32 F_\star^2 m_b^\star \pi^2 }  \frac{\left( 1 - \frac{m^2}{3 {m_b^\star}^2} \right)}{\sqrt{ 1 - \frac{ m^2}{4 {m_b^\star}^2}}}\arccos \left( - \frac{m^2}{2 {m_b^\star}^2} \right) \ .
\end{split}
\end{align}
\end{subequations}
$L$ contains the divergence and the finite constants typical for the modified minimal subtraction scheme in $4-\epsilon$ dimensions:%
\begin{align} \label{def_divergence}
L \equiv \frac{-1}{(4 \pi)^2} \left( \frac{1}{\epsilon}+\frac{1}{2} \left(1+\log{(4\pi)}-\gamma_E \right) \right) \ .
\end{align}
Note that we have shown that the divergencies of leading one-loop order can be canceled. For practical purposes one can therefore set $L$ to zero everywhere if one simultaneously replaces the corresponding DAs by the renormalized ones (compare section~\ref{sect_result}). Within our level of accuracy it is legitimate to replace $m_b^\star$ and $F_\star$ by their values at the symmetric point, where $\bar m_q = \bar m_q^\text{phys}$.%
\section{Handbook of distribution amplitudes}\label{app_handbook_of_DAs}
In this section we express the the $24$ standard DAs occurring in the general decomposition derived in ref.~\cite{Braun:2000kw} ($S_i^B$, $P_i^B$, $V_i^B$, $A_i^B$, $T_i^B$) in terms of the DAs defined in section~\ref{sect_result}. The equations given below follow directly from the definition of the DAs in eqs.~\eqref{definition_superior_DAs} and~\eqref{definition_superior_DAs_second_part} together with the symmetry properties of the standard DAs under exchange of the first and the second variable given in eq.~\eqref{symmetry_standard_DAs_12}. For the twist $3$ and twist $6$ amplitudes one finds%
\begin{align}
V^B_{1/6}&= \frac{1}{2}\biggl(\frac{\Phi^B_{+,3/6}{\scriptstyle(x_1,x_2,x_3)}}{c_B^+} + \frac{\Phi^B_{-,3/6}{\scriptstyle(x_1,x_2,x_3)}}{c_B^-} \biggr) + \frac{(-1)_B }{2} \biggl( \frac{\Phi^B_{+,3/6}{\scriptstyle(x_2,x_1,x_3)}}{c_B^+} + \frac{\Phi^B_{-,3/6}{\scriptstyle(x_2,x_1,x_3)}}{c_B^-}  \biggr)  \ , \nonumber \\
A^B_{1/6} &=   - \frac{1}{2}\biggl(\frac{\Phi^B_{+,3/6}{\scriptstyle(x_1,x_2,x_3)}}{c_B^+} + \frac{\Phi^B_{-,3/6}{\scriptstyle(x_1,x_2,x_3)}}{c_B^-} \biggr) + \frac{(-1)_B }{2} \biggl( \frac{\Phi^B_{+,3/6}{\scriptstyle(x_2,x_1,x_3)}}{c_B^+} + \frac{\Phi^B_{-,3/6}{\scriptstyle(x_2,x_1,x_3)}}{c_B^-}  \biggr) \ , \nonumber \\
T^B_{1/6} &= (-1)_B \frac{\Pi^B_{3/6}{\scriptstyle(x_1,x_3,x_2)}}{c_B^-} \ ,
\end{align}
where the DAs on the l.h.s.\ are functions of $(x_1,x_2,x_3)$. The twist $4$ and twist $5$ amplitudes read%
\begin{align}
S^B_{1/2} &= \frac{(-1)_B }{24}\biggl(\frac{\Xi^B_{+,4/5}{\scriptstyle(x_1,x_2,x_3)}}{c_B^+} + \frac{\Xi^B_{-,4/5}{\scriptstyle(x_1,x_2,x_3)}}{c_B^-} \biggr) - \frac{1 }{24}\biggl( \frac{\Xi^B_{+,4/5}{\scriptstyle(x_2,x_1,x_3)}}{c_B^+} + \frac{\Xi^B_{-,4/5}{\scriptstyle(x_2,x_1,x_3)}}{c_B^-}  \biggr) \nonumber \\*
&\quad + \frac{1}{4} \biggl(\frac{\Pi^B_{4/5}{\scriptstyle(x_2,x_3,x_1)}}{c_B^-} - (-1)_B \frac{\Pi^B_{4/5}{\scriptstyle(x_1,x_3,x_2)}}{c_B^-} \biggr) \ , \nonumber \\
P^B_{1/2} &= \frac{(-1)_B }{24}\biggl(\frac{\Xi^B_{+,4/5}{\scriptstyle(x_1,x_2,x_3)}}{c_B^+} + \frac{\Xi^B_{-,4/5}{\scriptstyle(x_1,x_2,x_3)}}{c_B^-} \biggr) - \frac{1 }{24}\biggl( \frac{\Xi^B_{+,4/5}{\scriptstyle(x_2,x_1,x_3)}}{c_B^+} + \frac{\Xi^B_{-,4/5}{\scriptstyle(x_2,x_1,x_3)}}{c_B^-}  \biggr) \nonumber \\*
&\quad - \frac{1}{4} \biggl(\frac{\Pi^B_{4/5}{\scriptstyle(x_2,x_3,x_1)}}{c_B^-} - (-1)_B \frac{\Pi^B_{4/5}{\scriptstyle(x_1,x_3,x_2)}}{c_B^-} \biggr) \ , \nonumber \\
V^B_{2/5} &= \frac{1}{4}\biggl(\frac{\Phi^B_{+,4/5}{\scriptstyle(x_1,x_2,x_3)}}{c_B^+} + \frac{\Phi^B_{-,4/5}{\scriptstyle(x_1,x_2,x_3)}}{c_B^-} \biggr) + \frac{(-1)_B }{4} \biggl( \frac{\Phi^B_{+,4/5}{\scriptstyle(x_2,x_1,x_3)}}{c_B^+} + \frac{\Phi^B_{-,4/5}{\scriptstyle(x_2,x_1,x_3)}}{c_B^-}  \biggr) \ , \nonumber \\
A^B_{2/5} &= - \frac{1}{4}\biggl(\frac{\Phi^B_{+,4/5}{\scriptstyle(x_1,x_2,x_3)}}{c_B^+} + \frac{\Phi^B_{-,4/5}{\scriptstyle(x_1,x_2,x_3)}}{c_B^-} \biggr) + \frac{(-1)_B }{4} \biggl( \frac{\Phi^B_{+,4/5}{\scriptstyle(x_2,x_1,x_3)}}{c_B^+} + \frac{\Phi^B_{-,4/5}{\scriptstyle(x_2,x_1,x_3)}}{c_B^-}  \biggr) \ , \nonumber \\
V^B_{3/4} &= \frac{(-1)_B}{4}\biggl(\frac{\Phi^B_{+,4/5}{\scriptstyle(x_3,x_1,x_2)}}{c_B^+} - \frac{\Phi^B_{-,4/5}{\scriptstyle(x_3,x_1,x_2)}}{c_B^-} \biggr) + \frac{1 }{4} \biggl( \frac{\Phi^B_{+,4/5}{\scriptstyle(x_3,x_2,x_1)}}{c_B^+} - \frac{\Phi^B_{-,4/5}{\scriptstyle(x_3,x_2,x_1)}}{c_B^-}  \biggr) \ , \nonumber \\
A^B_{3/4} &= - \frac{(-1)_B}{4}\biggl(\frac{\Phi^B_{+,4/5}{\scriptstyle(x_3,x_1,x_2)}}{c_B^+} - \frac{\Phi^B_{-,4/5}{\scriptstyle(x_3,x_1,x_2)}}{c_B^-} \biggr) + \frac{1 }{4} \biggl( \frac{\Phi^B_{+,4/5}{\scriptstyle(x_3,x_2,x_1)}}{c_B^+} - \frac{\Phi^B_{-,4/5}{\scriptstyle(x_3,x_2,x_1)}}{c_B^-}  \biggr) \ , \nonumber \\
T^B_{2/5} &= \frac{\Upsilon^B_{4/5}{\scriptstyle(x_3,x_2,x_1)}}{6 c_B^-} \ , \nonumber \\
T^B_{3/4} &= \frac{(-1)_B }{24}\biggl(\frac{\Xi^B_{+,4/5}{\scriptstyle(x_1,x_2,x_3)}}{c_B^+} + \frac{\Xi^B_{-,4/5}{\scriptstyle(x_1,x_2,x_3)}}{c_B^-} \biggr) + \frac{1 }{24}\biggl( \frac{\Xi^B_{+,4/5}{\scriptstyle(x_2,x_1,x_3)}}{c_B^+} + \frac{\Xi^B_{-,4/5}{\scriptstyle(x_2,x_1,x_3)}}{c_B^-}  \biggr) \nonumber \\*
&\quad + \frac{1}{4} \biggl(\frac{\Pi^B_{4/5}{\scriptstyle(x_2,x_3,x_1)}}{c_B^-} + (-1)_B \frac{\Pi^B_{4/5}{\scriptstyle(x_1,x_3,x_2)}}{c_B^-} \biggr) \ , \nonumber \\
T^B_{7/8} &= -\frac{(-1)_B }{24}\biggl(\frac{\Xi^B_{+,4/5}{\scriptstyle(x_1,x_2,x_3)}}{c_B^+} + \frac{\Xi^B_{-,4/5}{\scriptstyle(x_1,x_2,x_3)}}{c_B^-} \biggr) - \frac{1 }{24}\biggl( \frac{\Xi^B_{+,4/5}{\scriptstyle(x_2,x_1,x_3)}}{c_B^+} + \frac{\Xi^B_{-,4/5}{\scriptstyle(x_2,x_1,x_3)}}{c_B^-}  \biggr) \nonumber \\*
&\quad + \frac{1}{4} \biggl(\frac{\Pi^B_{4/5}{\scriptstyle(x_2,x_3,x_1)}}{c_B^-} + (-1)_B \frac{\Pi^B_{4/5}{\scriptstyle(x_1,x_3,x_2)}}{c_B^-} \biggr) \ .
\end{align}%
\section{Matching to other definitions in the literature}\label{app_matching}
Since we use the same definitions as ref.~\cite{Braun:2000kw} it is no surprise that%
\begin{align}
 f^{N} &= f_{N}(\text{\cite{Braun:2000kw}}) \ , &
 \lambda_1^{N} &= \lambda_{1}(\text{\cite{Braun:2000kw}}) \ , &
 \lambda_2^{N} &= \lambda_{2}(\text{\cite{Braun:2000kw}}) \ .
\end{align}
We can also match some of our constants to the leading twist normalization constants given in ref.~\cite{Chernyak:1987nu}. Note that for the $\Sigma$ and $\Xi$ a relative minus sign originates from the fact that ref.~\cite{Chernyak:1987nu} uses $\Sigma^+$ and $\Xi^-$ for the definition, while our choice is $\Sigma^-$ and $\Xi^0$ in order to have the same sign as for the proton.%
\begin{subequations}
\begin{align}
 f^{N} &= f_{N}(\text{\cite{Chernyak:1987nu}} ) \ , \\
 f^{\Sigma} &= -f_{\Sigma}(\text{\cite{Chernyak:1987nu}} ) \ , & 
 f_T^{\Sigma} &= -f^T_{\Sigma}(\text{\cite{Chernyak:1987nu}} ) \ ,\\
 f^{\Xi} &= -f_{\Xi}(\text{\cite{Chernyak:1987nu}} ) \ , & 
 f_T^{\Xi} &= -f^T_{\Xi}(\text{\cite{Chernyak:1987nu}} ) \ ,\\
 f^{\Lambda} &= \sqrt{\frac{2}{3}}f_{\Lambda}(\text{\cite{Chernyak:1987nu}} ) \ , & 
 \int [dx] x_1 \Phi^\Lambda_{-,3}(x_1,x_2,x_3) &= \sqrt{6}f^T_{\Lambda}(\text{\cite{Chernyak:1987nu}} ) \ .
\end{align}
\end{subequations}%
Due to some misprints, obvious errors and inconsistencies within refs.~\cite{Liu:2008yg,Liu:2009uc,Liu:2014uha} we are not able to give reliable matching formulas for their definitions.%
\section{Some construction details} \label{app_no_cov_der}
In the first part of this section we will describe why we can trade covariant derivatives acting on the baryon field for normal derivatives acting on the complete current. This choice is very convenient since the external derivatives (in contrast to the covariant derivatives acting on the baryon field) do not lead to additional loop momenta in the integrals. To show that this formulation only differs in higher order contributions we use the identities%
\begin{subequations}
\begin{align}
 \varepsilon^{abc}&=u_{aa^\prime} u_{bb^\prime} u_{cc^\prime} \varepsilon^{a^\prime b^\prime c^\prime} \ , \\ 
 \varepsilon^{abc}&=(u^\dagger)_{aa^\prime} (u^\dagger)_{bb^\prime} (u^\dagger)_{cc^\prime} \varepsilon^{a^\prime b^\prime c^\prime} \ , \\
 0&=\bigl( (\partial_\mu u)_{aa^\prime} u_{bb^\prime} u_{cc^\prime} +  u_{aa^\prime} (\partial_\mu u)_{bb^\prime} u_{cc^\prime} + u_{aa^\prime} u_{bb^\prime} (\partial_\mu u)_{cc^\prime} \bigr)  \varepsilon^{a^\prime b^\prime c^\prime} \ , \\
 0&=\bigl( (\partial_\mu u^\dagger)_{aa^\prime} (u^\dagger)_{bb^\prime} (u^\dagger)_{cc^\prime} +  (u^\dagger)_{aa^\prime} (\partial_\mu u^\dagger)_{bb^\prime} (u^\dagger)_{cc^\prime} + (u^\dagger)_{aa^\prime} (u^\dagger)_{bb^\prime} (\partial_\mu u^\dagger)_{cc^\prime} \bigr)  \varepsilon^{a^\prime b^\prime c^\prime} \ ,
\end{align}
\end{subequations}
which follow from $\det(u)=1$. From these one obtains%
\begin{align}
 \begin{split}
(D_\mu B)_{a a^\prime} \varepsilon^{a^\prime b c}  &= \bigl((\partial_\mu B)_{a a^\prime} \delta_{bb^\prime} \delta_{cc^\prime} + (\Gamma_\mu B)_{a a^\prime} \delta_{bb^\prime} \delta_{cc^\prime}  - ( B \Gamma_\mu)_{a a^\prime} \delta_{bb^\prime} \delta_{cc^\prime} \bigr) \varepsilon^{a^\prime b^\prime c^\prime}   \\
 &= \bigl((\partial_\mu B)_{a a^\prime} \delta_{bb^\prime} \delta_{cc^\prime} + (\Gamma_\mu B)_{a a^\prime} \delta_{bb^\prime} \delta_{cc^\prime}  + ( B )_{a a^\prime} (\Gamma_\mu)_{bb^\prime} \delta_{cc^\prime}\\*
&\quad + ( B)_{a a^\prime} \delta_{bb^\prime} (\Gamma_\mu)_{cc^\prime} \bigr) \varepsilon^{a^\prime b^\prime c^\prime} \ .
 \end{split}
\end{align}
Additionally we need%
\begin{subequations}
\begin{align}
\partial_\mu u_X &= u_X (u_{\bar X} \partial_\mu u_X) = u_X \Bigl( \Gamma_\mu -(-1)_X \frac{i}{2} u_\mu \Bigr) \mathrel{\dot =} u_X \Gamma_\mu \ , \\
\partial_\mu X_M &= D_\mu X_M - [\Gamma_\mu,X_M] \mathrel{\dot =} -[\Gamma_\mu,X_M] \ , \quad \text{for } X_M\in\{u_\nu,\chi_\pm\} \ .
\end{align}
\end{subequations}
Putting everything together we find for a general structure with arbitrary mesonic building blocks $X_1,X_2,X_3\in\{u_\nu,\chi_\pm\}$%
\begin{align}
 &(u_X X_1 D_\mu B)_{aa^\prime} (u_Y X_2)_{bb^\prime} (u_Z X_3)_{cc^\prime}  \varepsilon^{a^\prime b^\prime c^\prime} \nonumber\\
&\quad= \bigl(  (u_X X_1 \partial_\mu B)_{aa^\prime} (u_Y X_2)_{bb^\prime} (u_Z X_3)_{cc^\prime} +  (u_X X_1 \Gamma_\mu B)_{aa^\prime} (u_Y X_2)_{bb^\prime} (u_Z X_3)_{cc^\prime}  \nonumber \\
&\quad\quad +  (u_X X_1 B)_{aa^\prime} (u_Y X_2 \Gamma_\mu)_{bb^\prime} (u_Z X_3)_{cc^\prime}  +  (u_X X_1 B)_{aa^\prime} (u_Y X_2 )_{bb^\prime} (u_Z X_3 \Gamma_\mu)_{cc^\prime}   \bigr) \varepsilon^{a^\prime b^\prime c^\prime}  \nonumber \\
&\quad \mathrel{\dot =} \partial_\mu \bigl( (u_X X_1 B)_{aa^\prime} (u_Y X_2)_{bb^\prime} (u_Z X_3)_{cc^\prime} \bigr) \varepsilon^{a^\prime b^\prime c^\prime} \ .
\end{align}%
In the last step we have used%
\begin{align}
 u_X X_1 \Gamma_\mu \mathrel{\dot =} u_X \partial_\mu X_1 + u_X\Gamma_\mu X_1 \mathrel{\dot =}u_X \partial_\mu X_1 + (\partial_\mu u_X) X_1 = \partial_\mu ( u_X X_1)  \ ,
\end{align}%
and the same for $u_Y X_2$ and $u_Z X_3$.\par
In the following we will argue that structures involving baryon and meson fields at different positions can be dropped. We can choose the structure containing the baryon to be situated at $x$, while we call the second position $y$ such that we can write schematically $B(x,y) = f(x)g(y)$, where $g$ only contains mesonic building blocks. Every derivative acting on $g$ therefore has to be counted as first order in the chiral power counting. It follows trivially that%
\begin{align}
 f(x)g(y)&=f(x)\Bigl(g(x) + (x-y)\cdot\partial g(x) + \dots \bigr) \mathrel{\dot =} f(x)g(x) \ .
\end{align}%
\section{Matching relations}\label{app_projection_standard_DAs}
In this section we provide the result of the matching described in section~\ref{sect_matching}, which is needed in intermediate steps of our calculation (in practical applications one can always use the readily evaluated expressions shown in appendix~\ref{app_handbook_of_DAs}). For the twist projected amplitudes introduced in ref.~\cite{Braun:2000kw} one finds%
\begin{align}
S_1^B &=2 h^{2,abc}_{B,\cheven}{\scriptstyle(x_1,x_2,x_3)}+2 h^{2,abc}_{B,\chodd}{\scriptstyle(x_1,x_2,x_3)}\ , \nonumber \\
S_2^B &=-4 h^{1,abc}_{B,\cheven}{\scriptstyle(x_1,x_2,x_3)}+2 h^{2,abc}_{B,\cheven}{\scriptstyle(x_1,x_2,x_3)}-4 h^{1,abc}_{B,\chodd}{\scriptstyle(x_1,x_2,x_3)}+2 h^{2,abc}_{B,\chodd}{\scriptstyle(x_1,x_2,x_3)}\ , \nonumber \\
P_1^B &=2 h^{2,abc}_{B,\cheven}{\scriptstyle(x_1,x_2,x_3)}-2 h^{2,abc}_{B,\chodd}{\scriptstyle(x_1,x_2,x_3)}\ , \nonumber \\
P_2^B &=-4 h^{1,abc}_{B,\cheven}{\scriptstyle(x_1,x_2,x_3)}+2 h^{2,abc}_{B,\cheven}{\scriptstyle(x_1,x_2,x_3)}+4 h^{1,abc}_{B,\chodd}{\scriptstyle(x_1,x_2,x_3)}-2 h^{2,abc}_{B,\chodd}{\scriptstyle(x_1,x_2,x_3)}\ , \nonumber \\
V_1^B &=-4 h^{8,bca}_{B,\chodd}{\scriptstyle(x_2,x_3,x_1)}-4 h^{8,cab}_{B,\chodd}{\scriptstyle(x_3,x_1,x_2)}\ , \nonumber \\
V_2^B &=2 h^{2,bca}_{B,\chodd}{\scriptstyle(x_2,x_3,x_1)}+2 h^{7,bca}_{B,\chodd}{\scriptstyle(x_2,x_3,x_1)}-2 h^{8,bca}_{B,\chodd}{\scriptstyle(x_2,x_3,x_1)}\nonumber \\*&\quad-2 h^{2,cab}_{B,\chodd}{\scriptstyle(x_3,x_1,x_2)}+2 h^{7,cab}_{B,\chodd}{\scriptstyle(x_3,x_1,x_2)}  -2 h^{8,cab}_{B,\chodd}{\scriptstyle(x_3,x_1,x_2)}\ , \nonumber  \\
V_3^B &=-2 h^{2,bca}_{B,\chodd}{\scriptstyle(x_2,x_3,x_1)}+2 h^{7,bca}_{B,\chodd}{\scriptstyle(x_2,x_3,x_1)}-2 h^{8,bca}_{B,\chodd}{\scriptstyle(x_2,x_3,x_1)}\nonumber \\*&\quad+2 h^{2,cab}_{B,\chodd}{\scriptstyle(x_3,x_1,x_2)}+2 h^{7,cab}_{B,\chodd}{\scriptstyle(x_3,x_1,x_2)}  -2 h^{8,cab}_{B,\chodd}{\scriptstyle(x_3,x_1,x_2)}\ , \nonumber \\
V_4^B &=4 h^{1,bca}_{B,\chodd}{\scriptstyle(x_2,x_3,x_1)}-2 h^{2,bca}_{B,\chodd}{\scriptstyle(x_2,x_3,x_1)}+4 h^{4,bca}_{B,\chodd}{\scriptstyle(x_2,x_3,x_1)}+2 h^{7,bca}_{B,\chodd}{\scriptstyle(x_2,x_3,x_1)} \nonumber \\*&\quad-2 h^{8,bca}_{B,\chodd}{\scriptstyle(x_2,x_3,x_1)}  -4 h^{1,cab}_{B,\chodd}{\scriptstyle(x_3,x_1,x_2)}+2 h^{2,cab}_{B,\chodd}{\scriptstyle(x_3,x_1,x_2)}+4 h^{4,cab}_{B,\chodd}{\scriptstyle(x_3,x_1,x_2)} \nonumber \\*&\quad +2 h^{7,cab}_{B,\chodd}{\scriptstyle(x_3,x_1,x_2)}-2 h^{8,cab}_{B,\chodd}{\scriptstyle(x_3,x_1,x_2)}\ , \nonumber \\
V_5^B &=-4 h^{1,bca}_{B,\chodd}{\scriptstyle(x_2,x_3,x_1)}+2 h^{2,bca}_{B,\chodd}{\scriptstyle(x_2,x_3,x_1)}+4 h^{4,bca}_{B,\chodd}{\scriptstyle(x_2,x_3,x_1)}+2 h^{7,bca}_{B,\chodd}{\scriptstyle(x_2,x_3,x_1)} \nonumber \\*&\quad -2 h^{8,bca}_{B,\chodd}{\scriptstyle(x_2,x_3,x_1)}  +4 h^{1,cab}_{B,\chodd}{\scriptstyle(x_3,x_1,x_2)}-2 h^{2,cab}_{B,\chodd}{\scriptstyle(x_3,x_1,x_2)} +4 h^{4,cab}_{B,\chodd}{\scriptstyle(x_3,x_1,x_2)}\nonumber \\*&\quad +2 h^{7,cab}_{B,\chodd}{\scriptstyle(x_3,x_1,x_2)}-2 h^{8,cab}_{B,\chodd}{\scriptstyle(x_3,x_1,x_2)}\ , \nonumber \\
V_6^B &=8 h^{4,bca}_{B,\chodd}{\scriptstyle(x_2,x_3,x_1)}+8 h^{7,bca}_{B,\chodd}{\scriptstyle(x_2,x_3,x_1)}-4 h^{8,bca}_{B,\chodd}{\scriptstyle(x_2,x_3,x_1)}-16 h^{9,bca}_{B,\chodd}{\scriptstyle(x_2,x_3,x_1)}\nonumber \\*&\quad +8 h^{4,cab}_{B,\chodd}{\scriptstyle(x_3,x_1,x_2)} +8 h^{7,cab}_{B,\chodd}{\scriptstyle(x_3,x_1,x_2)}-4 h^{8,cab}_{B,\chodd}{\scriptstyle(x_3,x_1,x_2)}-16 h^{9,cab}_{B,\chodd}{\scriptstyle(x_3,x_1,x_2)}\ , \nonumber \\
A_1^B &=4 h^{8,cab}_{B,\chodd}{\scriptstyle(x_3,x_1,x_2)}-4 h^{8,bca}_{B,\chodd}{\scriptstyle(x_2,x_3,x_1)}\ , \nonumber \\
A_2^B &=2 h^{2,bca}_{B,\chodd}{\scriptstyle(x_2,x_3,x_1)}+2 h^{7,bca}_{B,\chodd}{\scriptstyle(x_2,x_3,x_1)}-2 h^{8,bca}_{B,\chodd}{\scriptstyle(x_2,x_3,x_1)}\nonumber \\*&\quad+2 h^{2,cab}_{B,\chodd}{\scriptstyle(x_3,x_1,x_2)}-2 h^{7,cab}_{B,\chodd}{\scriptstyle(x_3,x_1,x_2)}+2 h^{8,cab}_{B,\chodd}{\scriptstyle(x_3,x_1,x_2)}\ , \nonumber \\
A_3^B &=2 h^{2,bca}_{B,\chodd}{\scriptstyle(x_2,x_3,x_1)}-2 h^{7,bca}_{B,\chodd}{\scriptstyle(x_2,x_3,x_1)}+2 h^{8,bca}_{B,\chodd}{\scriptstyle(x_2,x_3,x_1)}\nonumber \\*&\quad+2 h^{2,cab}_{B,\chodd}{\scriptstyle(x_3,x_1,x_2)}+2 h^{7,cab}_{B,\chodd}{\scriptstyle(x_3,x_1,x_2)}-2 h^{8,cab}_{B,\chodd}{\scriptstyle(x_3,x_1,x_2)}\ , \nonumber \\
A_4^B &=-4 h^{1,bca}_{B,\chodd}{\scriptstyle(x_2,x_3,x_1)}+2 h^{2,bca}_{B,\chodd}{\scriptstyle(x_2,x_3,x_1)}-4 h^{4,bca}_{B,\chodd}{\scriptstyle(x_2,x_3,x_1)}-2 h^{7,bca}_{B,\chodd}{\scriptstyle(x_2,x_3,x_1)}\nonumber \\*&\quad+2 h^{8,bca}_{B,\chodd}{\scriptstyle(x_2,x_3,x_1)}-4 h^{1,cab}_{B,\chodd}{\scriptstyle(x_3,x_1,x_2)}+2 h^{2,cab}_{B,\chodd}{\scriptstyle(x_3,x_1,x_2)}+4 h^{4,cab}_{B,\chodd}{\scriptstyle(x_3,x_1,x_2)}\nonumber \\*&\quad+2 h^{7,cab}_{B,\chodd}{\scriptstyle(x_3,x_1,x_2)}-2 h^{8,cab}_{B,\chodd}{\scriptstyle(x_3,x_1,x_2)}\ , \nonumber \\
A_5^B &=-4 h^{1,bca}_{B,\chodd}{\scriptstyle(x_2,x_3,x_1)}+2 h^{2,bca}_{B,\chodd}{\scriptstyle(x_2,x_3,x_1)}+4 h^{4,bca}_{B,\chodd}{\scriptstyle(x_2,x_3,x_1)}+2 h^{7,bca}_{B,\chodd}{\scriptstyle(x_2,x_3,x_1)}\nonumber \\*&\quad-2 h^{8,bca}_{B,\chodd}{\scriptstyle(x_2,x_3,x_1)}-4 h^{1,cab}_{B,\chodd}{\scriptstyle(x_3,x_1,x_2)}+2 h^{2,cab}_{B,\chodd}{\scriptstyle(x_3,x_1,x_2)}-4 h^{4,cab}_{B,\chodd}{\scriptstyle(x_3,x_1,x_2)}\nonumber \\*&\quad-2 h^{7,cab}_{B,\chodd}{\scriptstyle(x_3,x_1,x_2)}+2 h^{8,cab}_{B,\chodd}{\scriptstyle(x_3,x_1,x_2)}\ , \nonumber \\
A_6^B &=8 h^{4,bca}_{B,\chodd}{\scriptstyle(x_2,x_3,x_1)}+8 h^{7,bca}_{B,\chodd}{\scriptstyle(x_2,x_3,x_1)}-4 h^{8,bca}_{B,\chodd}{\scriptstyle(x_2,x_3,x_1)}-16 h^{9,bca}_{B,\chodd}{\scriptstyle(x_2,x_3,x_1)}\nonumber \\*&\quad-8 h^{4,cab}_{B,\chodd}{\scriptstyle(x_3,x_1,x_2)}-8 h^{7,cab}_{B,\chodd}{\scriptstyle(x_3,x_1,x_2)}+4 h^{8,cab}_{B,\chodd}{\scriptstyle(x_3,x_1,x_2)}+16 h^{9,cab}_{B,\chodd}{\scriptstyle(x_3,x_1,x_2)}\ , \nonumber \\
T_1^B &=4 h^{8,abc}_{B,\chodd}{\scriptstyle(x_1,x_2,x_3)}\ , \nonumber \\
T_2^B &=-16 h^{5,abc}_{B,\cheven}{\scriptstyle(x_1,x_2,x_3)}\ , \nonumber \\
T_3^B &=-8 h^{5,abc}_{B,\cheven}{\scriptstyle(x_1,x_2,x_3)}+2 h^{6,abc}_{B,\cheven}{\scriptstyle(x_1,x_2,x_3)}-2 h^{7,abc}_{B,\chodd}{\scriptstyle(x_1,x_2,x_3)}+2 h^{8,abc}_{B,\chodd}{\scriptstyle(x_1,x_2,x_3)}\ , \nonumber \\
T_4^B &=-4 h^{3,abc}_{B,\cheven}{\scriptstyle(x_1,x_2,x_3)}-4 h^{4,abc}_{B,\cheven}{\scriptstyle(x_1,x_2,x_3)}-8 h^{5,abc}_{B,\cheven}{\scriptstyle(x_1,x_2,x_3)}-2 h^{6,abc}_{B,\cheven}{\scriptstyle(x_1,x_2,x_3)}\nonumber \\*&\quad-4 h^{4,abc}_{B,\chodd}{\scriptstyle(x_1,x_2,x_3)}-2 h^{7,abc}_{B,\chodd}{\scriptstyle(x_1,x_2,x_3)}+2 h^{8,abc}_{B,\chodd}{\scriptstyle(x_1,x_2,x_3)}\ , \nonumber \\
T_5^B &=-8 h^{4,abc}_{B,\cheven}{\scriptstyle(x_1,x_2,x_3)}-16 h^{5,abc}_{B,\cheven}{\scriptstyle(x_1,x_2,x_3)}\ , \nonumber \\ 
T_6^B &=-8 h^{4,abc}_{B,\chodd}{\scriptstyle(x_1,x_2,x_3)}-8 h^{7,abc}_{B,\chodd}{\scriptstyle(x_1,x_2,x_3)}+4 h^{8,abc}_{B,\chodd}{\scriptstyle(x_1,x_2,x_3)}+16 h^{9,abc}_{B,\chodd}{\scriptstyle(x_1,x_2,x_3)}\ , \nonumber \\
T_7^B &=8 h^{5,abc}_{B,\cheven}{\scriptstyle(x_1,x_2,x_3)}-2 h^{6,abc}_{B,\cheven}{\scriptstyle(x_1,x_2,x_3)}-2 h^{7,abc}_{B,\chodd}{\scriptstyle(x_1,x_2,x_3)}+2 h^{8,abc}_{B,\chodd}{\scriptstyle(x_1,x_2,x_3)}\ , \nonumber \\*
T_8^B &=4 h^{3,abc}_{B,\cheven}{\scriptstyle(x_1,x_2,x_3)}+4 h^{4,abc}_{B,\cheven}{\scriptstyle(x_1,x_2,x_3)}+8 h^{5,abc}_{B,\cheven}{\scriptstyle(x_1,x_2,x_3)}+2 h^{6,abc}_{B,\cheven}{\scriptstyle(x_1,x_2,x_3)}\nonumber \\*&\quad-4 h^{4,abc}_{B,\chodd}{\scriptstyle(x_1,x_2,x_3)} -2 h^{7,abc}_{B,\chodd}{\scriptstyle(x_1,x_2,x_3)}+2 h^{8,abc}_{B,\chodd}{\scriptstyle(x_1,x_2,x_3)}\ ,
\end{align}%
where the DAs on the l.h.s.\ are functions of $(x_1,x_2,x_3)$. The functions on the r.h.s.\ are given in eq.~\eqref{def_h_functions}. For the flavor indices $a,b,c$ on the r.h.s.\ one has to insert the flavors of the operators for which the l.h.s.\ is defined. A standard choice is $p \mathrel{ \hat = } uud$, $n \mathrel{ \hat = } ddu$, $\Sigma^+ \mathrel{ \hat = } uus$, $\Sigma^0  \mathrel{ \hat = } uds$, $\Sigma^- \mathrel{ \hat = } dds$, $\Xi^0 \mathrel{ \hat = } ssu$, $\Xi^- \mathrel{ \hat = } ssd$, $\Lambda \mathrel{ \hat = } uds$, where the order of the flavors is relevant for the symmetry properties of the DAs.\par%
\providecommand{\href}[2]{#2}\begingroup\raggedright\endgroup
\end{document}